\documentclass{mn2e}
\usepackage{graphicx}
\def\go{\mathrel{\raise.3ex\hbox{$>$}\mkern-14mu
             \lower0.6ex\hbox{$\sim$}}}
\def\lo{\mathrel{\raise.3ex\hbox{$<$}\mkern-14mu
             \lower0.6ex\hbox{$\sim$}}}
\def\msun{{\rm M_{\odot}}}
\def\etal{{et al.\ }}

\begin{document}

\title[{\it XMM-Newton} observations of high-luminosity radio-quiet QSOs]
{{\it XMM-Newton} observations of high-luminosity radio-quiet QSOs.}
\author[K.L. Page \etal]{K.L. Page$^{1}$,
J.N. Reeves$^{2}$, P.T. O'Brien$^{1}$, M.J.L. Turner$^{1}$ and D.M. Worrall$^{3}$\\
$^{1}$ X-Ray and Observational Astronomy Group, Department of Physics \& Astronomy,  
University of Leicester, LE1 7RH, UK\\
$^{2}$ Laboratory for High Energy Astrophysics, Code 662, NASA Goddard Space Flight Center, Greenbelt, MD 20771, USA\\
$^{3}$ H.H. Wills Physics Laboratory, University of Bristol, BS8 1TL, UK}

\date{Received ** *** 2004 / Accepted ** *** 2004}

\label{firstpage}

\maketitle

\begin{abstract}

{\it XMM-Newton} observations of five high-luminosity radio-quiet QSOs (Q~0144$-$3938, UM~269, PG~1634+706, SBS~0909+532 and PG~1247+267) are presented. Spectral energy distributions were calculated from the {\it XMM-Newton} EPIC (European Photon Imaging Camera) and OM (Optical Monitor) data, with bolometric luminosities estimated in the range from 7~$\times$~10$^{45}$ to 2~$\times$~10$^{48}$ erg s$^{-1}$ for the sample, peaking in the UV. At least four of the QSOs show a similar soft excess, which can be well modelled by either one or two blackbody components, in addition to the hard X-ray power-law. The temperatures of these blackbodies ($\sim$~100--500~eV) are too high to be direct thermal emission from the accretion disc, so Comptonization is suggested. Two populations of Comptonizing electrons, with different temperatures, are needed to model the broad-band spectrum. The hotter of these produces what is seen as the hard X-ray power-law, while the cooler ($\sim$~0.25--0.5~keV) population models the spectral curvature at low energies. Only one of the QSOs shows evidence for an absorption component, while three of the five show neutral iron emission. Of these, PG~1247+267 seems to have a broad line (EW~$\sim$~250~eV), with a strong, associated reflection component (R~$\sim$~2), measured out to 30~keV in the rest frame of the QSO.  Finally, it is concluded that the X-ray continuum shape of AGN remains essentially constant over a wide range of black hole mass and luminosity.


\end{abstract}


\section{Introduction}
\label{sec:intro}

\begin{figure*}
\begin{center}
\includegraphics[clip,height=11cm,angle=-90]{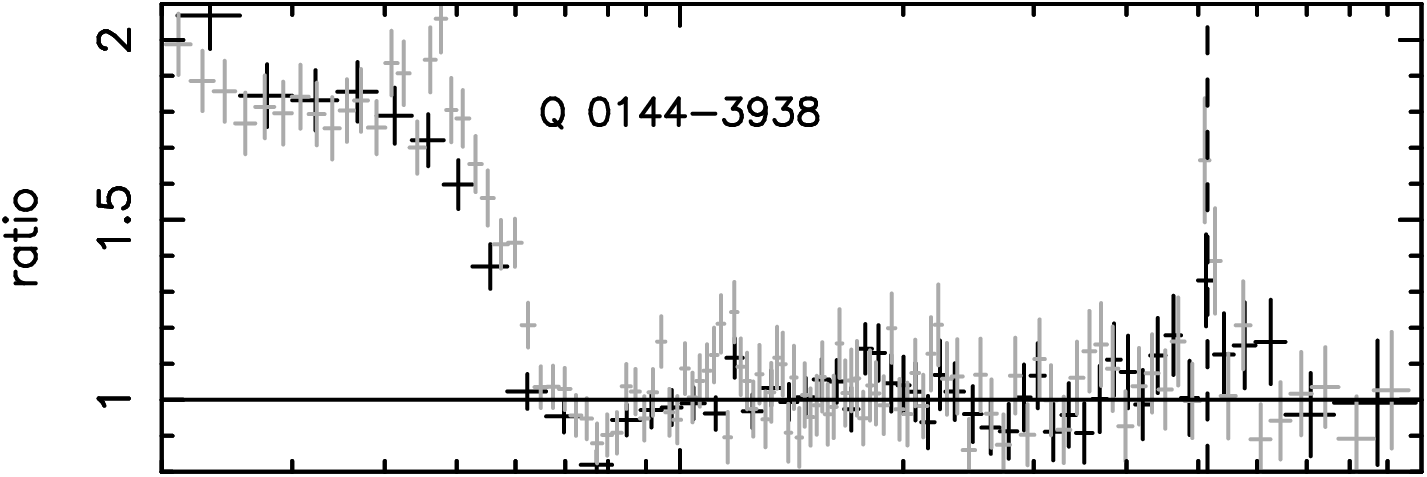}
\includegraphics[clip,height=11cm,angle=-90]{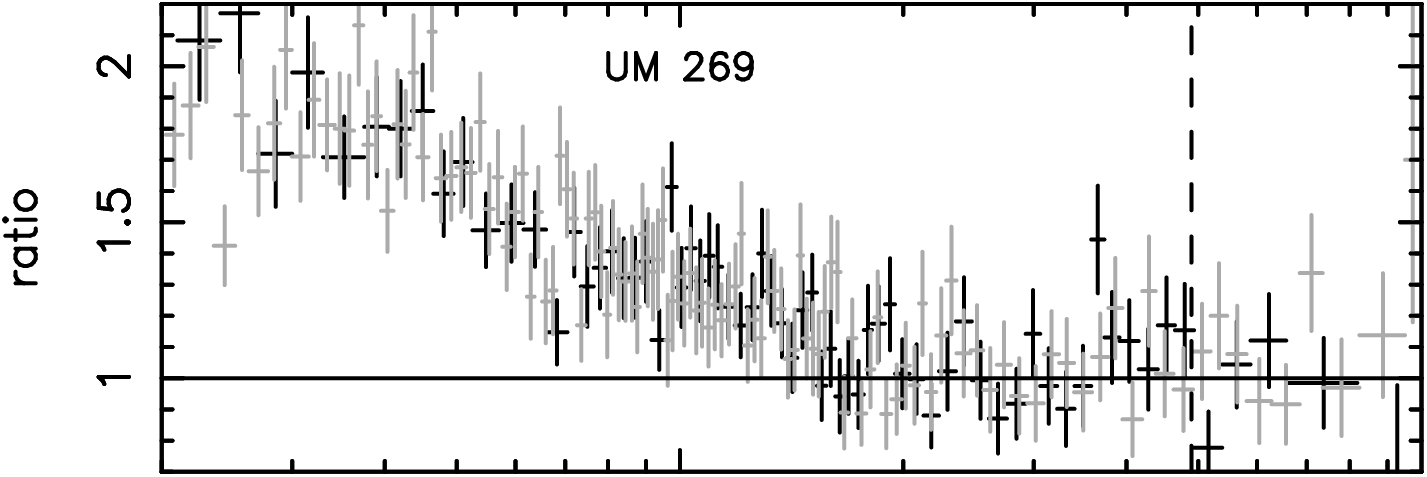}
\includegraphics[clip,height=11cm,angle=-90]{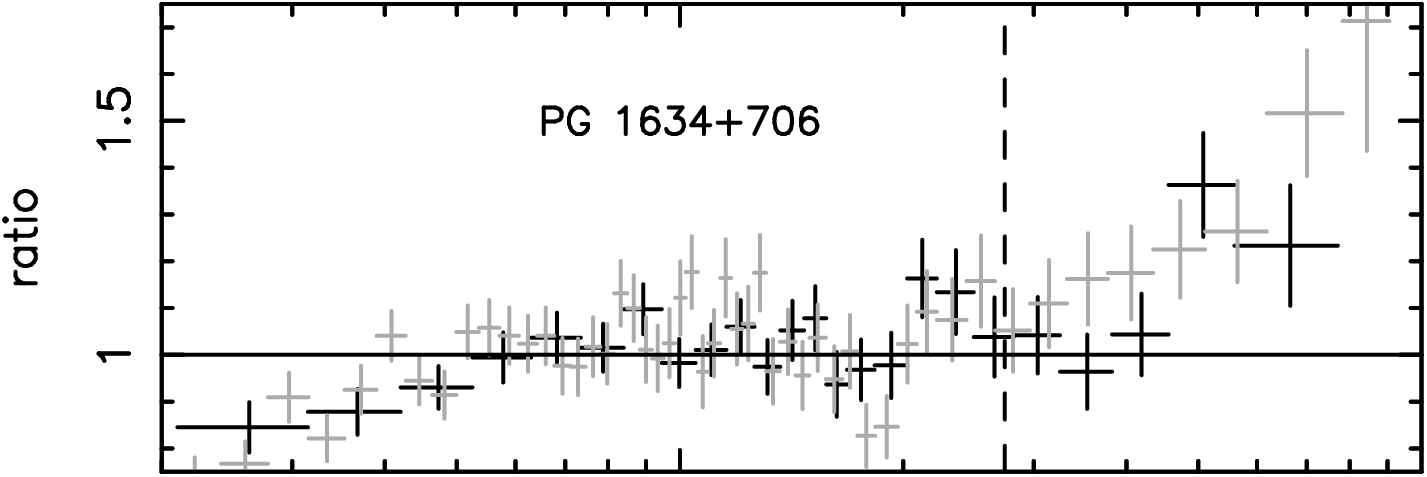}
\includegraphics[clip,height=11cm,angle=-90]{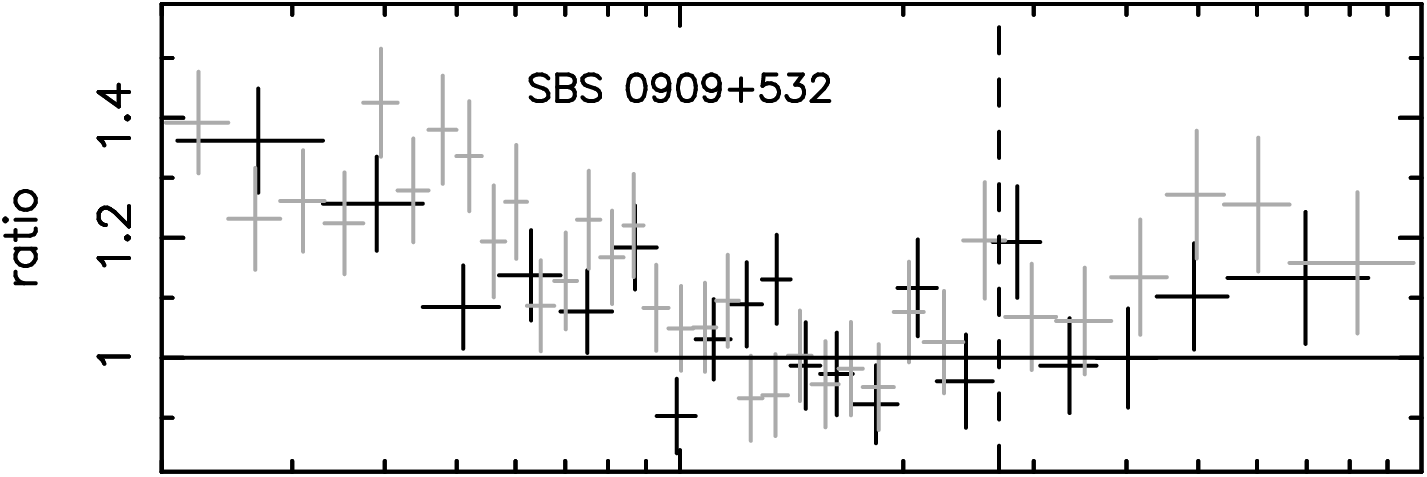}
\includegraphics[clip,height=11cm,angle=-90]{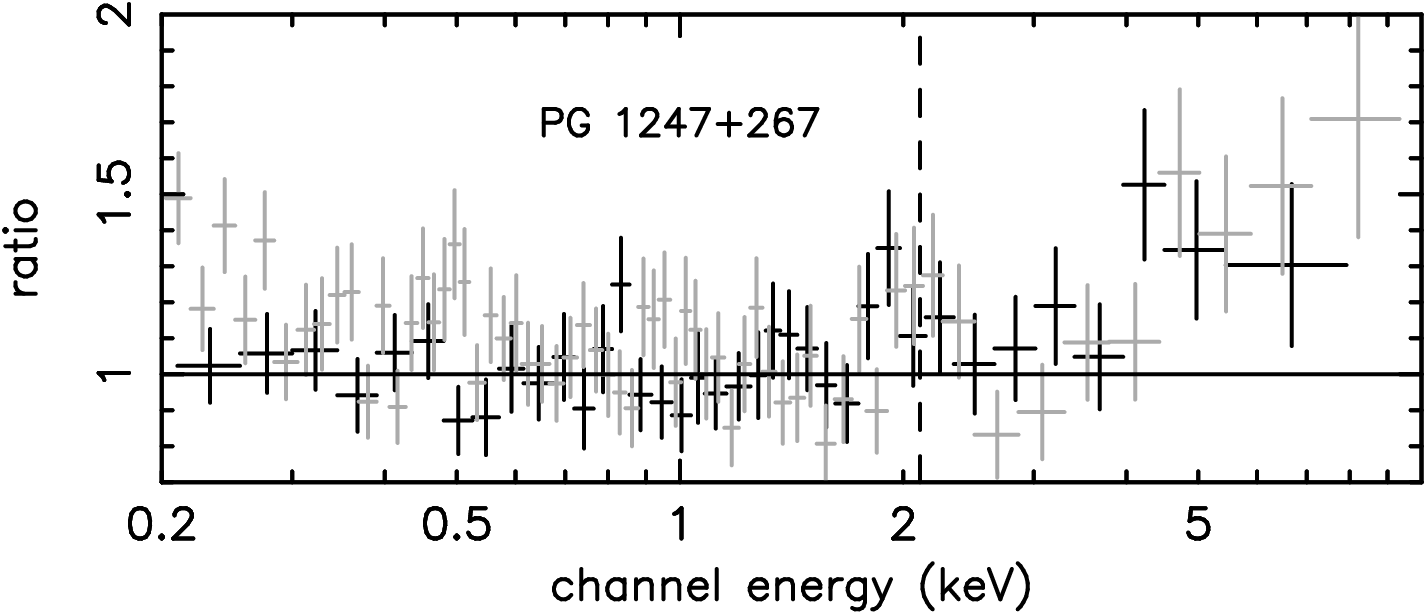}
\caption{The plots show the ratio of the data to a power-law model, fitted over 2--10~keV in the rest frame. Extrapolating the power-law fit over 0.2--10~keV (observer's frame) shows soft excesses in three of the objects; in the spectrum of PG~1247+267 the reflection component is dominant. The vertical dotted line marks the position of 6.4~keV in the rest frame. Positive residuals can be identified in the spectra of Q~0144$-$3938, PG~1247+267 and SBS~0909+532,  indicating the presence of iron emission. Co-added MOS residuals are shown in black, PN in grey.}
\label{se}
\end{center}
\end{figure*}

The X-ray continuum in Seyfert galaxies has been well studied; in these objects, a hard spectral component dominates the emission above $\sim$~2~keV. A proposed origin for this component is in a hot corona above the accretion disc surface, in which optical/UV photons from the disc are Comptonized to X-ray energies. A fraction of the hard X-ray power-law continuum then illuminates the disc (and possibly a molecular torus); some of these photons are absorbed, forming an iron K-edge at $>$~7~keV, while others are reprocessed into an Fe K$\alpha$ line at 6.4~keV and a Compton reflection `hump', caused by Compton down-scattering of the hard X-ray photons (Pounds \etal 1990).

It has been found that, although these features are generally observed in Seyfert 1 galaxies (Nandra \& Pounds 1994), they are less common in the spectra of QSOs (e.g., Reeves \etal 1997; Lawson \& Turner 1997). In many cases this could be related to a lack of signal to noise, but it is important to determine whether such features are ubiquitous in QSOs and, hence, whether this emission mechanism is common over the full range of Active Galactic Nuclei (AGN) luminosities. Observing the most luminous QSOs allows an investigation into objects where the accretion rate may be close to the Eddington limit and/or where the black hole mass may be large (i.e., $\sim$~10$^{9}$~$\msun$).

Work with {\it ASCA} (e.g., Reeves \& Turner 2000) found evidence for iron-line emission in a number of QSOs, but often originating from ionized material, rather than the cold emission found for Seyferts. If this were true in general, it could be explained by the more luminous AGN having a higher accretion rate, which causes the surface of the disc to become ionized.


The QSOs in this paper form a small sample of high-luminosity, radio-quiet objects, with X-ray luminosities from 7~$\times$~10$^{44}$ to 3~$\times$~10$^{46}$ erg~s$^{-1}$ (bolometric luminosities of 7~$\times$~10$^{45}$ to 2~$\times$~10$^{48}$ erg~s$^{-1}$); the redshifts cover a range from 0.244 up to 2.038. Radio-quiet objects form the majority of luminous AGN (e.g., Kukula \etal 1998), but are less X-ray luminous than their radio-loud counterparts, for a given optical luminosity (Zamorani \etal 1981; Worrall \etal 1987; Wilkes \etal 1994). Because radio-quiet QSOs (RQQs) are fainter in the X-ray band, the high through-put of the {\it XMM-Newton} X-ray telescopes makes the EPIC instruments ideal for an X-ray investigation of distant RQQs. The aim of this study is to investigate the properties of the central engine in some of the most luminous QSOs (i.e., the extreme end of the accretion rate, black hole mass and/or luminosity parameter space). It is possible to do this with RQQs since the jet is thought not to contribute significantly to the X-ray emission, whereas, in radio-loud quasars, Synchrotron or inverse Compton emission from a relativistic jet may dilute some of the spectral features (such as the iron line or soft excess) which are thought to originate from the accretion disc.

\section{XMM-Newton Observations}
\label{sec:xmmobs}

Table~\ref{xmm} lists the dates and the instrumental set-up for each of the EPIC observations, while Table~\ref{objects} gives the redshifts, Galactic absorbing column and radio measurements for each of the QSOs. A value of R$_{L}$~$<$~1 defines the AGN as being radio-quiet, where R$_{L}$ is given by the log of the ratio of the radio (5~GHz) to optical ({\it B}-band) fluxes (Wilkes \& Elvis 1987; Kellerman \etal 1989). 

The pipeline-produced event-lists were filtered using {\sc xmmselect} within version 5.4 of the {\sc sas} (Science Analysis Software); single- and double-pixel events (patterns 0--4) were used for the PN, while patterns 0--12 were chosen for the MOS instruments. Spectra were extracted within a small circular region, centred on the source, with a radius of between 25 and 40 arcsec, depending on the source brightness. (Smaller regions were used for the fainter sources, to minimise the contribution from the background.)  Background spectra (within the same size, or larger, region) were produced from an off-set position free of other sources. MOS 1 and MOS 2 spectra were subsequently co-added, after confirmation that the results were consistent. Source and background light-curves were also extracted for each object. Time intervals of relatively high, flaring background were identified for both of the PG QSOs, and these periods were excluded from the following analysis. None of the source light-curves showed variability over the duration of the observations after the removal of the background, however. After grouping the spectra to obtain a minimum of 20 counts per bin, version 11.1.0 of {\sc Xspec} was used to analyse the data. The most recent (time-dependent) response matrices (rmfs) were used when fitting the spectra, together with an ancillary response file (arf) generated by running {\sc arfgen} within the {\sc sas}. The rmfs take into account the degradation of the instruments over the years since launch, and specifically model how the Charge Transfer Inefficiency (CTI) has changed. Optical/UV magnitudes were obtained from the Optical Monitor where possible and these are discussed in Section~\ref{sec:om}. Errors are given
at the 90~per~cent level (e.g., $\Delta\chi^{2}$~=~4.6 for 2 interesting parameters). Throughout this paper, a WMAP Cosmology of H$_{0}$~=~70~km~s$^{-1}$~Mpc$^{-1}$, Cosmological constant $\Omega_{\lambda}$~=~0.73 and $\Omega_{m}$~=~1-$\Omega_{\lambda}$ is assumed; for comparison with previous conventions, using q$_{0}$~=~0.1 and H$_{0}$~=~70~km~s$^{-1}$~Mpc$^{-1}$ would give slightly smaller luminosity distances: a factor of $\sim$0.93--0.97 (z~=~0.224--2.038) compared to the values used here.

\begin{table*}
\begin{center}
\caption{Information about the observations and {\it XMM-Newton} configurations for the QSOs. $^{a}$ This first observation of SBS 0909+532 was wiped out due to high background; $^{b}$ `clean' exposure times are given, after excluding periods of high background flares; $^{c}$ lw -- large window, ff -- full-frame}

\label{xmm}
\begin{tabular}{p{2.0truecm}p{1.5truecm}p{1.5truecm}p{0.9truecm}p{0.9truecm}p{0.9truecm}p{1.5truecm}p{0.9truecm}p{0.9truecm}p{0.9truecm}}
\hline
Object & Obs. ID & Obs. date & \multicolumn{3}{c} {exposure time (ks)$^{b}$} & filter & \multicolumn{3}{c} {mode$^{c}$}\\
 &  & (rev.) & MOS 1 & MOS 2 & PN & MOS/PN& MOS 1 & MOS 2 & PN \\
\hline
Q~0144$-$3938 & 0090070101 & 2002-06-18 (0462) & 32.3 & 32.3 & 28.1 & medium & lw & lw & ff\\
UM~269 & 0090070201 & 2002-01-05 (0380) &20.3 & 20.3 & 16.3 &  medium & lw & lw & ff\\
PG~1634+706 & 0143150101 & 2002-11-22 (0541) & 13.7 & 13.7 & 12.6  & medium & ff & ff & ff\\
SBS~0909+532$^{a}$ & 0143150301 & 2003-04-17 (0614) & 12.9 & 12.9 & 10.7 & medium & ff & ff & ff\\
& 0143150601 & 2003-05-18 (0630) & 17.0 & 17.1  & 13.8 & medium  & ff & ff & ff\\
PG~1247+267 & 0143150201 & 2003-06-18 (0645) & 24.0 & 24.0 & 19.0 & medium & ff & ff & ff\\

\hline
\end{tabular}
\end{center}
\end{table*}

\begin{table*}
\begin{center}
\caption{Information about the five QSOs in this sample. E(B$-$V) values taken from NED. The radio fluxes were measured at 1.5 GHz.
(i) NRAO/VLA Sky Survey (Condon \etal 1998); (ii) Wadadekar \& Kembhavi (1999); (iii) Vignali \etal (1999); (iv) Barvainis, Lonsdale \& Antonucci (1996); (v) Reeves \& Turner (2000); (vi) FIRST survey (Becker, White \& Helfand 1995); (vii) Kellerman \etal (1989)} 
\label{objects}
\begin{tabular}{p{2.0truecm}p{1.5truecm}p{1.5truecm}p{1.5truecm}p{2.0truecm}p{1.0truecm}p{1.8truecm}p{0.9truecm}p{2.0truecm}}
\hline
Object & RA & dec. & redshift &  Galactic N$_{H}$ &  E(B$-$V) & radio flux & R$_{L}$ & reference\\
&J2000 &J2000 & & (10$^{20}$ cm$^{-2}$) & & density (mJy)\\
\hline
Q~0144$-$3938 & 01:46:12.5 & $-$39:23:5.0 & 0.244 & 1.44 & 0.014 & 2.5  & $<$0.2 & (i)\\
UM~269 & 00:43:19.7 & 00:51:15.0 & 0.308 & 2.30 & 0.02 & 1.1  & 0.67 & (ii), (iii)\\
PG~1634+706 & 16:34:28.9 & 70:31:33 & 1.334 & 4.48 & 0.04 & 1.65 & $-$0.59 & (iv), (v)\\
SBS~0909+532 & 09:13:1.6 & 52:59:29.1 & 1.376 & 1.6 & 0.015 & 1.09 & $-$0.92 & (vi)\\
PG~1247+267 & 12:50:5.6 & 26:31:9.8 & 2.038 & 0.9 & 0.013 & 1.17  & 0.36 & (vii)\\
\hline
\end{tabular}
\end{center}
\end{table*}

\section{Analysis of EPIC data}
\label{sec:specanal}
\subsection{The iron K band}

Throughout the analysis, the co-added MOS and PN spectra were found to give consistent results, so the values given in this paper are those for joint fits to the instruments. A simple power-law fit over the 0.2--10~keV band showed significant curvature in all five spectra (Figure~\ref{se}). For three of the QSOs, upward curvature was found at low ($<$1--2~keV observer's frame) energies, indicating soft excess emission; although the panel for PG~1634+706 does not show an obvious excess of counts at lower energies, it was found to be better fitted by the inclusion of a soft excess component (see Section~\ref{broad}). The last panel of Figure~\ref{se} (PG~1247+267) shows an excess of counts in the high-energy portion of the spectrum. To avoid the curvature, the spectra over the rest-frame 2--10~keV bands were initially investigated. First, a power-law model, attenuated by neutral absorption, was tried, where N$_{H}$ was fixed at the Galactic value for each object, calculated using the {\bf nh} {\sc ftool} (Dickey \& Lockman 1990). Next a Gaussian component was included for each spectrum, to measure emission from Fe K$\alpha$; if such a line was insignificant, then only the upper limit on the equivalent width is quoted in Table~\ref{hard}. As the table shows, two of the objects did not require any iron emission components. The remaining three QSOs, however, did show significant evidence for neutral Fe K emission, at $>$~99~per~cent confidence, according to an F-test. 


Allowing the width of the line in Q~0144$-$3938 to vary improved the fit slightly over a narrow line model (null probability of 1.7~$\times$~10$^{-2}$); the best-fit line was found to have an intrinsic width of $\sigma$~$\sim$~0.15~keV. The spectrum of SBS~0909+532 also showed evidence for a line, but it was not possible statistically to differentiate between a narrow and a broadened component; hence, the 90~per~cent upper limit to the width is given in Table~\ref{hard}. Finally, PG~1247+267 showed iron emission as well. This line was also better modelled with a broad component, with an EW of $\sim$~400~eV for a width of $\sigma$~$\sim$~0.52~keV (decrease in $\chi^{2}$/dof of 5/1, compared to a narrow line; the F-test null probability for this is 8.9~$\times$~10$^{-3}$). It should be noted that Protassov \etal (2002) have shown that the F-test may not be an appropriate statistic for determining the significance of marginal line parameters and should be used with caution. Thus, although there is evidence that the iron line may be broadened, no definite statement can be made.

None of the five QSO spectra showed any evidence for iron absorption edges at $\go$7~keV, in contrast to PDS~456, for example (Reeves, O'Brien \& Ward 2003), which is also a high-luminosity radio-quiet AGN.

\begin{figure}
\begin{center}
\includegraphics[clip,width=5.0cm,angle=-90]{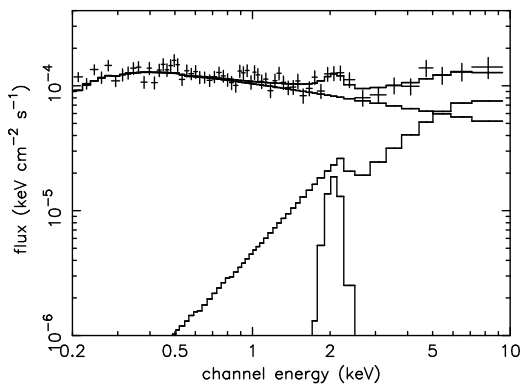}
\caption{The best fit to the broad-band spectrum of PG~1247+267 includes a strong reflection component, as this unfolded plot shows. There is also a correspondingly strong, broad Fe K$\alpha$ emission line. PN data only are shown for clarity.}
\label{refl}
\end{center}
\end{figure}

\begin{table*}
\begin{center}
\caption{Fits over the 2--10~keV energy band; 90~per~cent errors (Q~0144$-$3938, SBS~0909+532 and PG~1247+267) and  upper limits (UM~269 and  PG~1634+706) are given for the EW of the iron line. The width of the lines have been corrected to the rest-frame of the objects in question. The F-test null probability compares the fit with the iron line to that with only the power-law, with the number of extra free parameters in brackets. .} 
\label{hard}
\begin{tabular}{p{1.6truecm}p{0.3truecm}p{1.1truecm}p{1.8truecm}p{2.0truecm}p{1.5truecm}p{1.5truecm}p{1.1truecm}p{2.6truecm}}
\hline
Object & Fit & Model & $\Gamma$ & Line & $\sigma$ & Equivalent & $\chi^{2}$/dof & F-test null  \\
 & & & &  Energy (keV) & (keV) & Width (eV) & & prob. (free param.)\\
\hline

Q~0144$-$3938 & 1 & PL & 1.78~$\pm$~0.05 & & &  &343/369\\
& 2 & PL+GA & 1.82~$\pm$~0.05 & 6.45~$\pm$~0.06 & 0.15~$\pm$~0.1 & 215~$\pm$~80 &  317/366  & 2.36$\times$10$^{-6}$ (3) \\
UM~269 & 1 & PL & 1.68~$\pm$~0.10 & & & &211/198 \\
 & 2 & PL+GA & 1.69~$\pm$~0.10 & 6.4$^{f}$ & 0.01$^{f}$ & $<$80 & 210/197  & 0.334 (1) \\
PG~1634+706 & 1 & PL & 2.19~$\pm$~0.05 & & & &285/296\\
 & 2 & PL+GA & 2.19~$\pm$~0.05 & 6.4$^{f}$ & 0.01$^{f}$ & $<$72 & 284/295  & 0.309 (1) \\
SBS~0909+532 & 1 & PL & 1.71~$\pm$~0.05 & & && 205/227\\
 & 2 & PL+GA & 1.73~$\pm$~0.06 & 6.42~$\pm$~0.11 & $<$0.49 & 200~$\pm$~75 & 196/224  & 6.41$\times$10$^{-3}$ (3) \\
PG~1247+267 & 1 & PL & 2.15~$\pm$~0.07 & & && 123/160\\
 & 2 & PL+GA & 2.23~$\pm$~0.10 & 6.30~$\pm$~0.36 & 0.52~$\pm$~0.35 & 421~$\pm$~215 & 112/157 & 2.04$\times$10$^{-3}$ (3) \\
\hline
\end{tabular}
\end{center}
\end{table*}

\subsection{Broad-band X-ray continuum}
\label{broad}

Since three of the QSOs in this sample are at z~$>$~1, the presence of a 
Compton reflection hump was investigated, using 
the model {\it pexrav} (Magdziarz \& Zdziarski 1995) in {\sc Xspec}. In 
the standard model of AGN, reflection can occur 
from the accretion disc and/or molecular torus (e.g., Lightman \& White 
1988; George \& Fabian 1991). At z~=~1, the 
{\it XMM-Newton} band extends up to 20~keV in the rest frame, so it might 
be expected that any spectral flattening due to 
reflection would be observable. Since reflection components can also affect the lower-energy spectra, these fits were conducted over the full 0.2--10~keV energy band. The results of this model, given in Table~\ref{reftab}, show that only in PG~1247+267 was a reflection component significantly detected; upper 
limits are given for the remaining four objects. For PG~1247+267, a simple power-law plus Gaussian fit over the broad-band gave a reduced $\chi^{2}$ value of 270/295; the inclusion of the reflection parameter decreased this to 236/294, giving a null probability of 3.3~$\times$~10$^{-10}$. The component in PG~1247+267 is strong, with $R$~$\sim$~2, and is required 
in addition to the broad line; $R$ is defined to 
be $\Omega$/2$\pi$, where $\Omega$ is the solid angle subtended by the 
scattering medium. Thus, for a value of $R$~$>$~1, 
the indication is that reflection is occuring from $>$2$\pi$ steradian, 
which is unphysical; possible explanations are 
discussed in Section~\ref{disc}. 

\begin{figure}
\begin{center}
\includegraphics[clip,width=5.0cm,angle=-90]{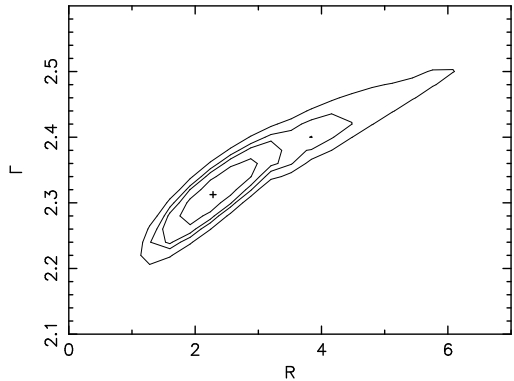}
\caption{68, 90, 95 and 99 per~cent confidence contours (two interesting parameters) for the strength of the reflection parameter in PG~1247+267 against photon index. The data are not consistent with zero reflection.}
\label{cont}
\end{center}
\end{figure}

As one would expect the strength of the Compton reflection hump to scale 
directly with the equivalent width of the iron 
line for a given photon index and disc inclination, the iron 
line equivalent width was linked to the strength of the 
reflection component, $R$, within the {\it pexrav} model. To determine the 
scaling between $R$ and the line equivalent 
width, the model of George \& Fabian (1991) was used, appropriate for 
reflection off neutral matter. In the case where the 
continuum photon index is $\Gamma$~$\sim$~2.3, with an inclination angle for 
the reflector of 30 degrees to the line of sight, 
a reflection component with $R$~=~1 should produce an iron line with an equivalent width of $\sim$~120~eV. Refitting this 
reflection model to PG~1247+267 then resulted in a 
slightly weaker iron line, with an equivalent width of 259~$\pm$~74~eV, 
whilst the strength of the corresponding 
reflection component is then $R$~=~2.2~$\pm$~0.7, consistent with the previous 
fit. 
The energy and velocity width of the line are largely unchanged, with 
E~=~6.35~$\pm$~0.51~keV and 
$\sigma$~=~0.53~$\pm$~0.37~keV. The reflection component is clearly detected 
in Figure~\ref{se}, with a null probability of 5.4~$\times$~10$^{-10}$ (for the presence of both the line and reflection component); this corresponds to $\Delta\chi^2$~=~40 for three degrees of freedom. The reflection hump is consistent with 
the strong line observed and  
the fit with both the broad line and reflector is shown in 
Figure~\ref{refl}. Figure~\ref{cont} shows the confidence 
contours for the strength of the reflection and the power-law 
slope, demonstrating clearly that the data are inconsistent with a 
reflection parameter of zero at $>$99~per~cent.

\begin{table}
\begin{center}
\caption{Compton reflection fits to the data. The reflection component, modelled using {\it pexrav}, is given by $R$~=~$\Omega$/2$\pi$. These values were obtained by modelling the broad-band (0.2--10~keV) X-ray spectra of the objects. F-test probabilities are given for one additional degree of freedom.} 
\label{reftab}
\begin{tabular}{p{2.0truecm}p{3.0truecm}p{2.2truecm}}
\hline
Object & Reflection parameter & F-test null prob.\\
\hline

Q~0144$-$3938 & $<$1.33 & 0.341\\
UM~269 & $<$2.00 & 0.999\\
PG~1634+706 & $<$0.72 & 0.567\\
SBS~0909+532 &  $<$1.1 & 0.999\\
PG~1247+267 & 2.87~$\pm$~0.96 & 3.3~$\times$~10$^{-10}$\\
\hline
\end{tabular}
\end{center}
\end{table}

\begin{figure}
\begin{center}
\includegraphics[clip,width=5.0cm,angle=-90]{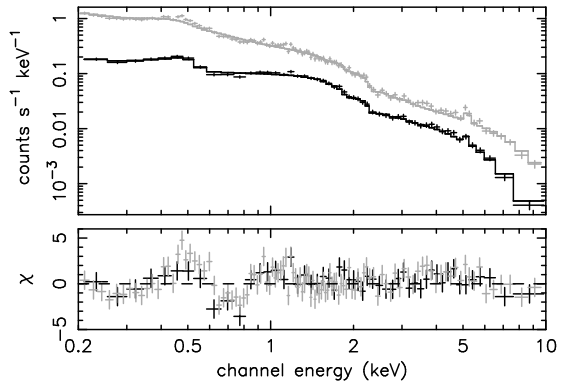}
\caption{After modelling the broad-band spectrum of Q~0144$-$3938 with a power-law, two blackbodies and an iron line, a dip can be seen in the spectrum around 0.6--0.8~keV (observer's frame); this indicates the presence of the warm absorber. MOS data are plotted in black, PN in grey.}
\label{wa}
\end{center}
\end{figure}

Extrapolating the best fit 2--10~keV models over the full {\it XMM-Newton} bandpass showed obvious soft excesses (Figure~\ref{se}) in three of the five QSOs; fit 3 in Table~\ref{broadband} gives the reduced $\chi^{2}$ value for this extrapolation. The soft excess emission appear strongest in those QSOs at the lower redshifts; however, this is mainly a selection effect, since the soft excess will be shifted out of the {\it XMM-Newton} energy bandpass as the redshift increases. Likewise, at higher z, any reflection component in the spectrum will be shifted into the hard band, making high-energy curvature more obvious in the data. The most straightforward, if not particularly physical, method for modelling the soft excess is the inclusion of blackbody (BB) components. As Table~\ref{broadband} shows, the objects were generally best fitted with one or two BBs in addition to the power-law (fits 5 and 6); however, the fit to Q~0144$-$3938 improves enormously with the addition of a warm absorber component (modelled with  {\it absori} in {\sc Xspec}), in the rest-frame of the AGN. The presence of this component was suggested in Figure~\ref{wa} by a dip in the spectrum around 0.6--0.8~keV. The main component of this warm absorber are edges corresponding to O{\sc vii} and O{\sc viii}. The other QSOs showed no evidence for excess N$_{H}$ and upper limits for cold absorption are given in Table~\ref{broadband}. It was found that there was only weak evidence for a soft emission component in PG~1247+267, with the reflection component accounting for most of the spectral curvature. Alternatively, the spectral shape of PG~1247+267 can be modelled by a much flatter power-law, together with three blackbody components and the broad line, but no Compton reflection hump. For $\Gamma$~=~1.61~$\pm$~0.14 and blackbody temperatures of kT~=~0.067~$\pm$~0.014, 0.267~$\pm$~0.045 and 0.702~$\pm$~0.089~keV, a $\chi^{2}$ value of 229 for 288 degrees of freedom is obtained.  However, this model is inconsistent with the presence of the strong emission line, which is an indication that a reflection component should be present (assuming the reprocessing medium is Compton-thick).

An alternative method for modelling the curvature is using a broken power-law to model the entire spectrum; these fits are quoted in Table~\ref{bkn}. For four of the QSOs, the break energies for the broken power-law (given in the rest-frame) are 1--4~keV, indicative of modelling soft excess emission; PG~1247+267 shows spectral hardening at higher energies, which is more suggestive of reflection, agreeing with the previous fits. Only the spectrum of SBS~0909+532 can be fitted statistically as well with a broken power-law as with the power-law/blackbody model. However, for the case of PG~1634+706, including an additional rest-frame absorption component [N$_{H}$~=~(16~$\pm$~2)~$\times$~10$^{20}$ cm$^{-2}$] actually improves the fit, despite there being no evidence for such absorption when modelling the spectrum with a power-law and blackbody. Blackbody spectra curve over at lower energies, while power-laws don't; Figure~\ref{se} shows that the spectrum of PG~1634+706 does roll over slightly at the low energies, hence the requirement for the excess absorption. This fit results in $\chi^{2}$/dof~=~487/506, a decrease of seven for one degree of freedom, compared to the blackbody fit. UM~269, SBS~0909+532 and PG~1247+267 still show no sign of an increased column density.


Table~\ref{lum} lists the 2--10 and 0.2--10~keV fluxes and unabsorbed luminosities, derived from the best-fit power-law/blackbody models. Note that SBS~0909+532 is a gravitational lens (Kochanek \etal 1997; Oscoz \etal 1997; Leh{\' a}r \etal 2002); hence, the luminosity estimated from the fits is much too high. Chartas \etal (2000) estimate that the luminosity of this object is enhanced by a factor of fifteen, thus the values determined from fitting the X-ray spectrum have been scaled down by this amount in the tables and figures in this paper.




Although blackbodies parametrize the soft excess very well, the temperatures for these components are significantly hotter than could be feasibly expected from a luminous QSO accretion disc. There is a relationship between the disc temperature, accretion rate and the mass of the black hole, given by (Peterson 1997):

\begin{equation}
T(r) \sim
6.3\times10^{5}\left(\frac{\dot{M}}{\dot{M}_{Edd}}\right)^{1/4}M_{8}^{-1/4}\left(\frac{r}{R_{sch}}\right)^{-3/4} \mathrm{K}
\label{tempeqn}
\end{equation}

\noindent For a black hole mass of $\sim$~10$^{9}$~$\msun$, the hottest temperature (i.e., at an innermost radius of 3~R$_{sch}$, where the Schwarzschild radius R$_{sch}$~=~2GM/c$^{2}$) is $\sim$~10~eV. The temperatures found by fitting the soft excess with blackbody components (Table~\ref{broadband}) are much higher than this, generally with kT~$\go$~100~eV. This indicates that the excess seen is not due to direct thermal emission from the accretion disc and requires Compton up-scattering of the disc photons.

\begin{table*}
\begin{center}
\caption{Fits over the 0.2--10~keV energy bands; the F-test values are given for each blackbody model compared to the previous fit. The Fe lines present in Q~0144$-$3938, SBS 0909+267 and PG 1247+267 were fixed to the values determined over the 2--10~keV band. The `EXTRAP. PL' fit refers to the $\chi^{2}$ value obtained if the power-law fitted to the 2--10~keV rest frame is simply extrapolated over the full 0.2--10~keV band. The fits for PG 1247+267 also include a reflection component, REF (see Table~\ref{reftab}).
$^{a}$ blackbody temperature; $^{b}$ absorption component; $^{c}$ ionization of the warm absorber; $^{d}$ This soft emission component is only marginally significant.} 
\label{broadband}
\begin{tabular}{p{1.6truecm}p{0.3truecm}p{2.5truecm}p{1.5truecm}p{1.8truecm}p{1.8truecm}p{1.7truecm}p{0.9truecm}p{1.2truecm}p{1.4truecm}}
\hline
Object & Fit & Model & $\Gamma$ & kT$^{a}$ & kT$^{a}$ & N$_{H}^{b}$ & $\xi^{c}$ &$\chi^{2}$/dof & F-test null\\
 & & & &  (keV) & (keV) & (10$^{20}$ cm$^{-2}$) & & & probability\\
\hline

Q~0144$-$3938 & 3 & EXTRAP. PL & 1.82~$\pm$~0.05 & & & & & 3289/754\\
& 4 & PL & 2.07~$\pm$~0.05 & & & & & 2112/754\\
 & 5 & PL+BB & 1.74~$\pm$~0.04 & 0.095~$\pm$~0.003 & & & & 956/752 & 1.00~$\times$~10$^{-99}$\\
 & 6 & PL+2BB & 1.54~$\pm$~0.07 & 0.103~$\pm$~0.003 & 0.488~$\pm$~0.030 & & & 902/750 & 3.39~$\times$~10$^{-10}$\\
 & 7 & (PL+2BB)*ABS & 1.81~$\pm$~0.05 & 0.123~$\pm$~0.008 & 0.283~$\pm$~0.020 & 97~$\pm$~29 & 98~$\pm$~32 & 828/748 & 1.25~$\times$~10$^{-14}$\\

UM~269 & 3 & EXTRAP. PL & 1.68~$\pm$~0.10 & & & & & 1824/514\\
& 4 & PL & 1.93~$\pm$~0.03 & & & & & 621/514\\
& 5 & PL+BB & 1.84~$\pm$~0.04 & 0.120~$\pm$~0.016 & & & & 564/512 & 1.97~$\times$~10$^{-11}$\\
 & 6 & PL+2BB & 1.59~$\pm$~0.10 & 0.104~$\pm$~0.011 & 0.311~$\pm$~0.035 & $<$8 & & 533/510 & 5.49~$\times$~10$^{-7}$\\

PG~1634+706 & 3 & EXTRAP. PL & 2.19~$\pm$~0.05 & & & & & 606/509\\
& 4 & PL & 2.11~$\pm$~0.02 & & & & &  548/509\\
& 5 & PL+BB & 1.96~$\pm$~0.05 & 0.404~$\pm$~0.027 & &$<$7 & & 494/507 & 3.79~$\times$~10$^{-12}$\\


SBS~0909+532 & 3 & EXTRAP. PL & 1.73~$\pm$~0.06 & & & & & 514/428\\
& 4 & PL & 1.81~$\pm$~0.01 & & & & &  424/428\\
& 5 & PL+BB & 1.64~$\pm$~0.05 & 0.274~$\pm$~0.034 & & &  & 375/426 & 4.36~$\times$~10$^{-12}$\\
 & 6 & PL+2BB$^{d}$ & 1.60~$\pm$~0.04 & 0.102~$\pm$~0.039 & 0.335~$\pm$~0.038 & $<$1 & & 367/424 & 1.00~$\times$~10$^{-2}$\\

PG~1247+267 & 3 & EXTRAP. PL & 2.23~$\pm$~0.10 & & & & & 287/295\\
& 4 & PL & 2.18~$\pm$~0.02 & & & & & 270/295\\
& 4.5 & PL+REF & 2.34~$\pm$~0.04 & & & & & 236/293 & 2.73~$\times$~10$^{-9}$\\
 &5 &  PL+REF+BB$^{d}$ & 2.30~$\pm$~0.05 & 0.050~$\pm$~0.020 & & $<$9 & &  230/291 & 2.00~$\times$~10$^{-2}$\\

\hline
\end{tabular}
\end{center}
\end{table*}

\begin{table*}
\begin{center}
\caption{Broken power-law fits to the broad-band spectra; absorption/Gaussian components are included as before. $\Gamma_{soft}$ is the photon index below the break energy (E$_{break}$), $\Gamma_{hard}$ above. The break energy is given in the QSO rest frames. The final column gives the change in $\chi^{2}$/degrees of freedom between the broken power-law fits and power-law/blackbody model; a negative sign demonstrates that the $\chi^{2}$ value is lower for the blackbody fit.}
$^{a}$ See text for an alternative fit, including absorption. 
\label{bkn}
\begin{tabular}{p{2.0truecm}p{0.6truecm}p{1.0truecm}p{2.0truecm}p{2.0truecm}p{2.0truecm}p{2.0truecm}p{2.0truecm}}
\hline
Object & Fit & Model & $\Gamma_{soft}$ & E$_{break}$ & $\Gamma_{hard}$  & $\chi^{2}$/dof & $\Delta\chi^{2}$/dof\\
\hline

Q~0144$-$3938 & 8 & BKNPL & 2.20~$\pm$~0.03 & 3.35~$\pm$~0.76 & 1.82~$\pm$~0.06 & 848/750 & $-$20/2\\ 
UM~269 & 8 & BKNPL & 2.07~$\pm$~0.03 & 2.68~$\pm$~0.43 & 1.65~$\pm$~0.08 & 544/512 & $-$11/2\\
PG~1634+706$^{a}$ & 8 & BKNPL & 2.12~$\pm$~0.01 & 1.63~$\pm$~0.38 & 1.84~$\pm$~0.17 & 541/507 & $-$47/1\\
SBS~0909+532 & 8 & BKNPL &  1.95~$\pm$~0.05 & 3.80~$\pm$~0.81 & 1.58~$\pm$~0.08 & 368/426 & $-$1/2\\
PG~1247+267 & 8 & BKNPL & 2.24~$\pm$~0.04 & 7.93~$\pm$~1.73 & 1.63~$\pm$~0.24 & 239/293 & $-$9/2\\
\hline
\end{tabular}
\end{center}
\end{table*}

\begin{table*}
\begin{center}
\caption{Fluxes and luminosities over the 2--10~keV (rest frame) and 0.2--10~keV (observer's frame) energy bands, calculated from the power-law plus blackbody models. The luminosities of SBS~0909+532 have been decreased by a factor of fifteen due to the object being a gravitational lens; the fluxes, however, are the actual observed values. The sixth column gives the rest-frame energy range to which the {\it XMM-Newton} band of 0.2--10~keV corresponds. The bolometric luminosities (estimated from 2500\AA) and (lower-limit) black hole masses are also quoted.} 
\label{lum}
\begin{tabular}{p{1.8truecm}p{2.0truecm}p{1.5truecm}p{2.0truecm}p{1.5truecm}p{2.5truecm}p{1.5truecm}p{1.5truecm}}
\hline
 & \multicolumn{2}{c} {2--10~keV (rest-frame)} & \multicolumn{2}{c} {0.2--10~keV (observer's frame)} \\
Object & flux & luminosity & flux & luminosity & rest frame & bolometric & black hole\\
 & (erg cm$^{-2}$~s$^{-1}$) & (erg~s$^{-1}$) & (erg cm$^{-2}$~s$^{-1}$)  & (erg~s$^{-1}$) & {\it XMM-Newton} band & luminosity (erg~s$^{-1}$) & mass ($\msun$)\\
\hline
Q~0144$-$3938  & 1.21~$\times$~10$^{-12}$ & 2.32~$\times$~10$^{44}$ & 2.58~$\times$~10$^{-12}$ & 6.60~$\times$~10$^{44}$ & 0.25--12.44 & 2~$\times$~10$^{46}$ & 2~$\times$~10$^{8}$\\
UM~269 & 8.90~$\times$~10$^{-13}$ & 2.80~$\times$~10$^{44}$ &  1.92~$\times$~10$^{-12}$ & 6.54~$\times$~10$^{44}$ & 0.26--13.08 & 7~$\times$~10$^{45}$  & 5~$\times$~10$^{7}$\\
PG~1634+706 & 1.16~$\times$~10$^{-12}$ & 1.32~$\times$~10$^{46}$ & 2.98~$\times$~10$^{-12}$ & 3.30~$\times$~10$^{46}$ & 0.47--23.34 & 2~$\times$~10$^{48}$ & 1~$\times$~10$^{10}$\\
SBS~0909+532 & 3.70~$\times$~10$^{-13}$ & 5.18~$\times$~10$^{44}$ & 1.53~$\times$~10$^{-12}$ & 1.29~$\times$~10$^{45}$ & 0.48--23.76 & 4~$\times$~10$^{46}$ & 3~$\times$~10$^{8}$\\
PG~1247+267 & 6.00~$\times$~10$^{-14}$ & 9.21~$\times$~10$^{45}$ & 7.00~$\times$~10$^{-13}$ & 2.78~$\times$~10$^{46}$ & 0.61--30.38 & 1~$\times$~10$^{48}$ & 1~$\times$~10$^{10}$\\
\hline
\end{tabular}
\end{center}
\end{table*}

Inverse Comptonization of the thermal photons can form both the soft excess and the apparent hard X-ray power-law. This theory suggests that there is a population of hot electrons, above the disc, which Comptonizes the disc photons to produce the observed soft excess. The `power-law' at higher energies could also be formed by Comptonization, but through interactions with a much hotter electron corona. Comptonized spectra `roll over' at $\sim$~4kT; hence, for a relatively cool ($<$~1~keV, say) electron population, the component will be similar in shape to a BB, only somewhat broader. For a hot electron population ($\sim$100~keV), the spectrum will appear power-law-like out to energies well beyond the energy bandpass of most current X-ray instruments. 

\begin{figure*}
\begin{center}
\includegraphics[clip,width=5.0cm,angle=-90]{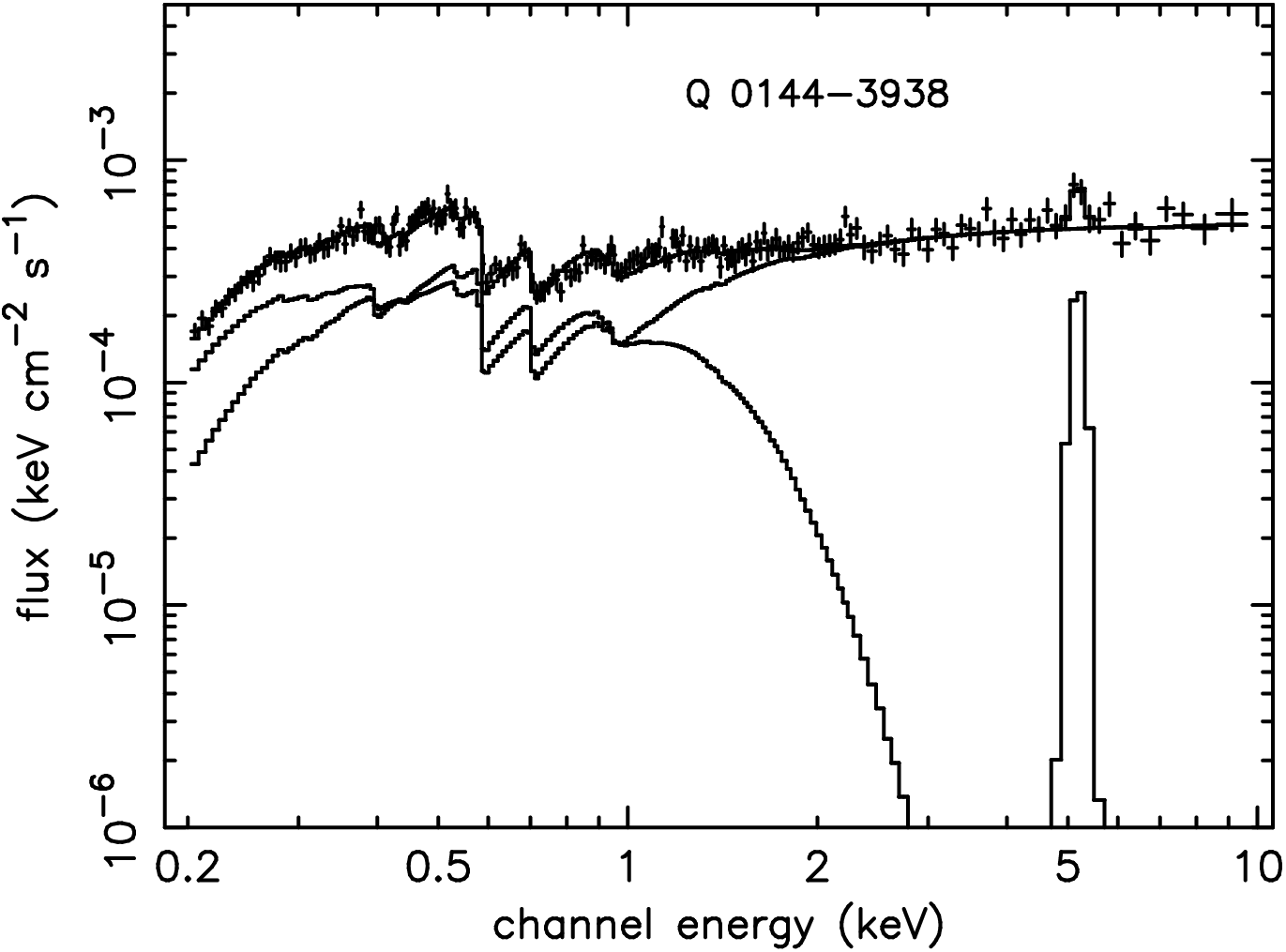}\hspace*{0.5cm}
\includegraphics[clip,width=5.0cm,angle=-90]{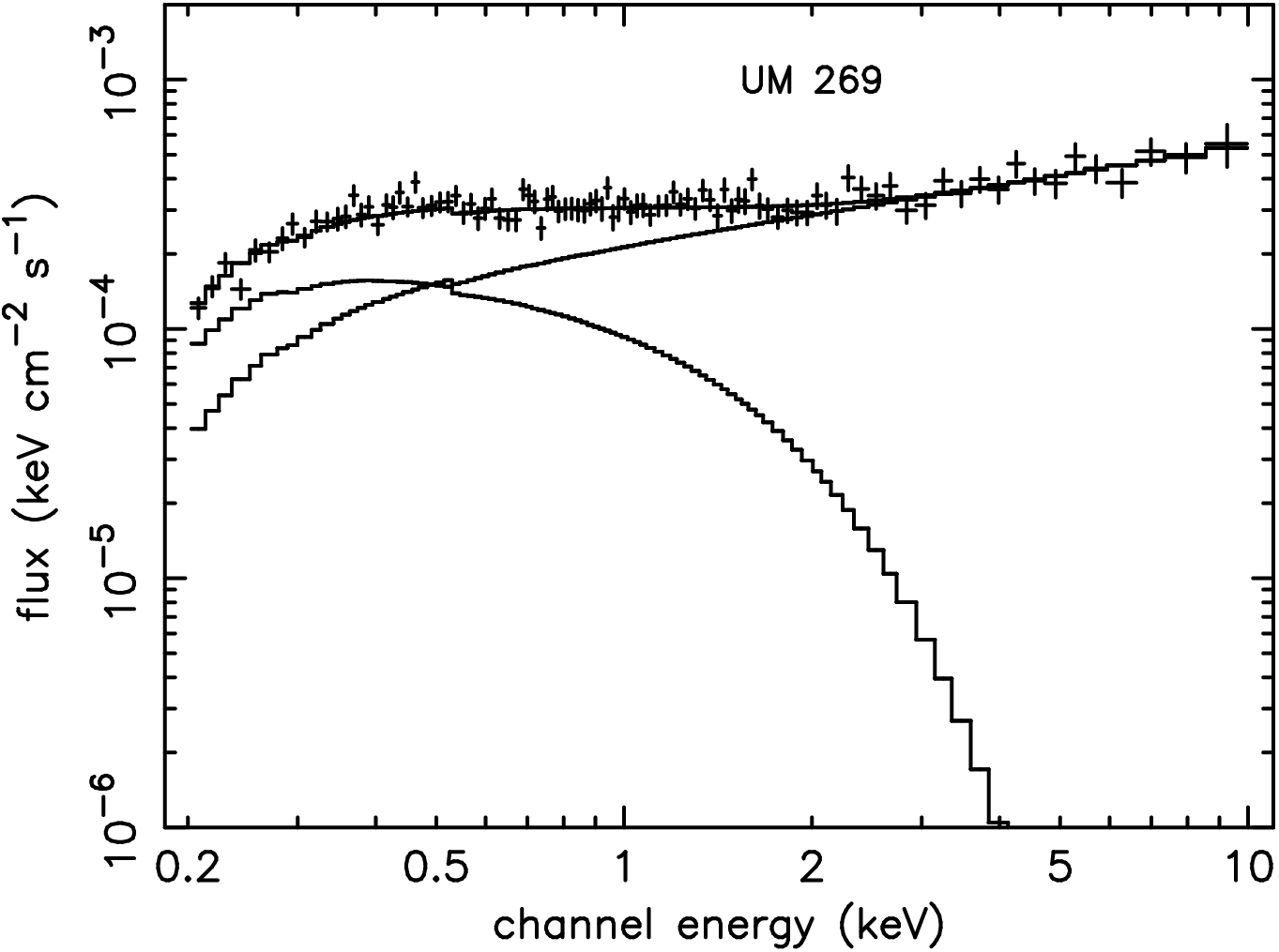}\vspace*{0.7cm}
\includegraphics[clip,width=5.0cm,angle=-90]{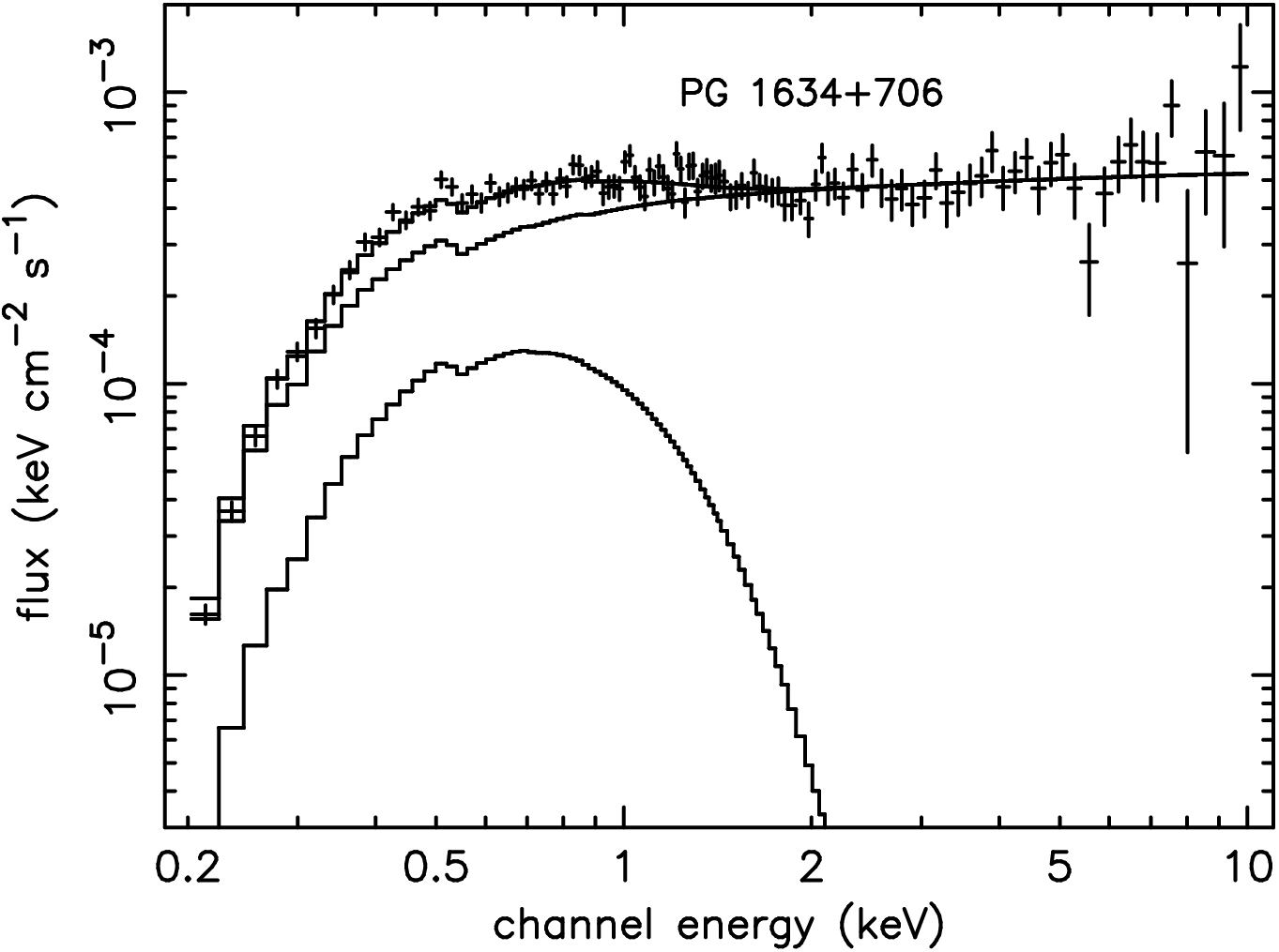}\hspace*{0.5cm}
\includegraphics[clip,width=5.0cm,angle=-90]{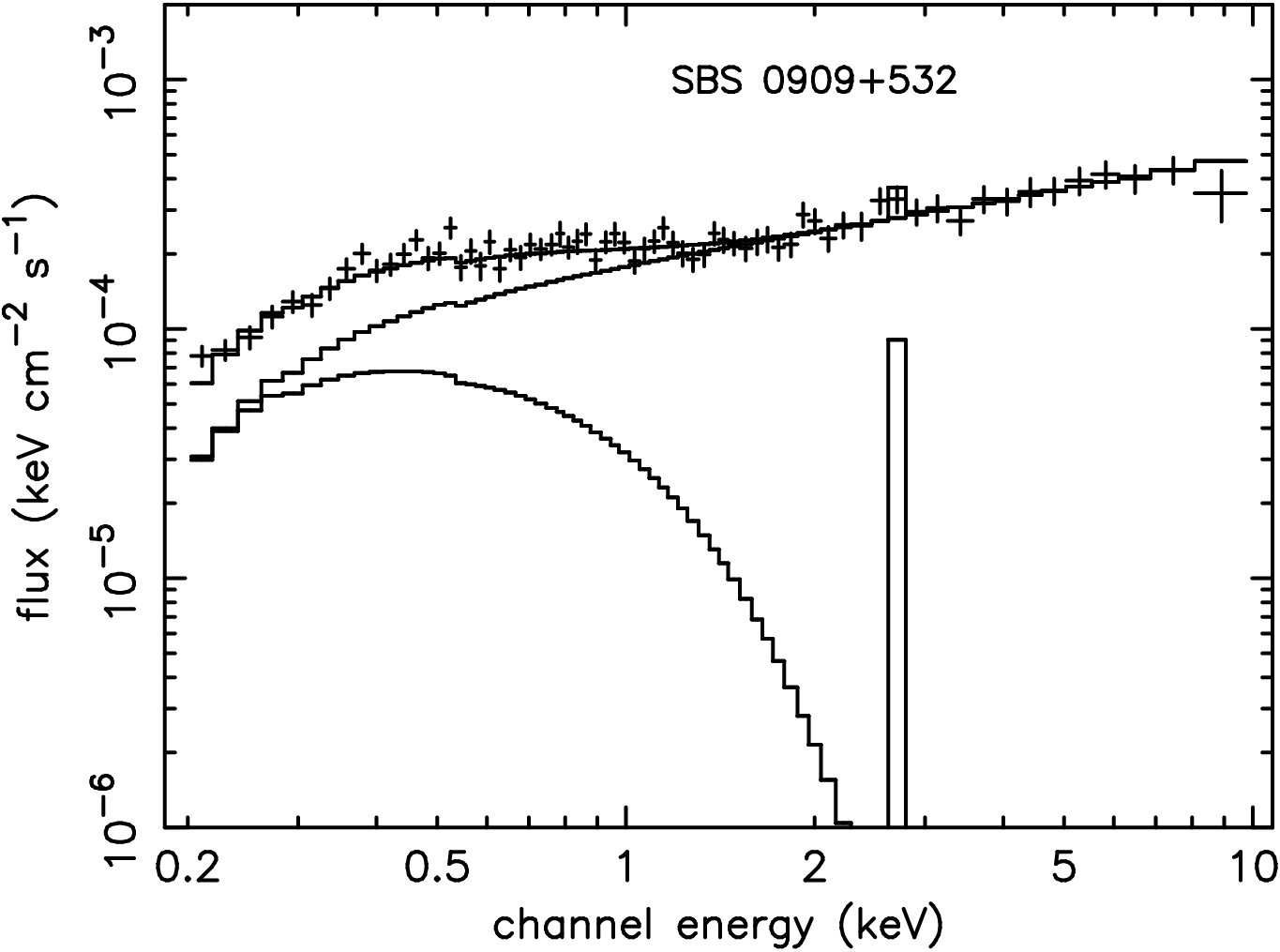}\vspace*{0.7cm}
\caption{Unfolded plots showing the Comptonization fits to the spectra of Q~0144$-$3938, UM~269, PG~1634+706, SBS~0909+532 and PG~1247+267. As before, Q~0144$-$3938 includes additional absorption.}
\label{thcomp_euf}  
\end{center}
\end{figure*}

The Comptonization model used to fit these spectra was {\it thCompfe} (Zdziarski, Johnson \& Magdziarz 1996); the results are shown in Table~\ref{compt} and the fits in Figure~\ref{thcomp_euf}. The temperature of the hottest Comptonizing distribution was set to 200~keV, since the roll-off for the `power-law' component is expected to be outside the bandpass of {\it XMM-Newton}, as explained above. This component will appear very similar to a power-law over this limited energy band. Modelling the spectra with an actual power-law (with a Comptonization component for the soft excess) leads to almost identical fits; Comptonization of the disc photons by hot electrons is, however, a plausible explanation for the formation of the observed power-law emission. The model for Q~0144$-$3938 includes a similar value for excess absorption as for the blackbody fit. The temperature of the disc photons will be cooler than can easily be modelled over the {\it XMM-Newton} band. Subsequently, an estimate of the accretion disc temperature was made as follows: the luminosity at 2500\AA\ (rest frame) was estimated from the $V$-band magnitude assuming an slope of 0.7; from Elvis \etal (1994), a typical QSO has a median bolometric to 2500\AA\ luminosity ratio of L$_{Bol}$/L$_{2500\AA}$ of 5.6. This allowed an estimate of the black hole mass to be obtained (Table~\ref{lum}), since 

\begin{equation}
L_{Edd} = \frac{4 \pi GM m_{p} c}{\sigma_{T}}
\approx 1.3 \times 10^{38} \frac{M}{\msun} {\mathrm{erg s^{-1}}}
\end{equation}

\noindent This estimate will be a lower-limit to the mass, since it has been assumed that accretion occurs at the maximum Eddington rate. Taking an accretion rate of $\dot{M}_{Edd}$ to be a typical value for high luminosity QSOs, the temperature of the disc was calculated for each object, for a radius of 3R$_{sch}$, using Equation~\ref{tempeqn}. The resultant temperatures were rounded to the nearest 5~eV, and are listed in the fourth column of Table~\ref{compt}. 

Two comments must be made about these Comptonization fits.  Firstly, the resulting model was largely insensitive to the estimated disc temperatures, because they are so low. Secondly, the geometry of AGN is unknown: it could be that some disc photons are Comptonized to soft excess energies, while others directly interact with the hotter electron population, forming the hard power-law. Alternatively, the two Comptonizing electron populations could be `layered', such that some photons are first Comptonized by the warm, `soft excess' population, before undergoing further Comptonization with the hotter electron cloud (due to the disc corona). If this were to be the case, then the input temperatures for the two components should not be the same. The results given in Table~\ref{compt} assume the first of these possible scenarios. The fits were also tried allowing the input photons to the hotter Comptonizing electrons to have a varying temperature; however, statistically the fits could not be differentiated for these spectra. Since it was found earlier that the spectrum of PG~1247+267 could be well described without a soft excess component if reflection were included, these data were not modelled with the Comptonization components.

\begin{table*}
\begin{center}
\caption{Double-Comptonization fits to the data. The absorption component was included for Q~0144$-$3938. $^{a}$ input blackbody temperature; $^{b}$ electron temperature; $^{c}$ optical depth; $^{f}$ frozen} 
\label{compt}
\begin{tabular}{p{1.9truecm}p{0.3truecm}p{1.8truecm}p{1.0truecm}p{1.8truecm}p{1.5truecm}p{1.0truecm}p{1.0truecm}p{1.5truecm}p{1.2truecm}}
\hline
&  &  &  \multicolumn{3}{c} {\sc cooler comptonized component} & \multicolumn{3}{c} {\sc hotter comptonized component}\\
Object & Fit & Model & kT$_{bb}^{a,f}$ & kT$_{e}^{b}$ & $\tau^{c}$ & kT$_{bb}^{a,f}$ & kT$_{e}^{b,f}$ & $\Gamma$ & $\chi^{2}$/dof\\
 & & & (eV) & (keV) & & (eV) & (keV)\\
\hline
Q~0144$-$3938 & 9 & 2{\sc thCompfe} &  20 & 0.261~$\pm$~0.016 & 32~$\pm$~4 & 20 & 200 & 1.90~$\pm$~0.04 & 836/749 \\

&\\
UM~269 & 9 & 2{\sc thCompfe} &  30 & 0.496~$\pm$~0.168 & 13~$\pm$~3 & 30 & 200 & 1.62~$\pm$~0.09  & 544/512\\

&\\
PG~1634+706 & 9 & 2{\sc thCompfe} & 10 & 0.420~$\pm$~0.060 & 40~$\pm$~18 & 10 & 200 & 1.95~$\pm$~0.05 & 541/506\\

&\\
SBS~0909+532 & 9 & 2{\sc thCompfe} & 10 & 0.491~$\pm$~0.169 & 15~$\pm$~5 & 10 & 200 & 1.60~$\pm$~0.05 & 368/425\\

&\\

\hline
\end{tabular}
\end{center}
\end{table*}

\section{OM data}
\label{sec:om}

With the exception of UM 269, OM observations were obtained for the AGN in this sample; Table~\ref{omdata} lists the magnitudes found in each band. Also given is $\alpha_{ox}$, the two-point optical to X-ray spectral index. As for estimating the bolometric luminosity, a typical optical slope of 0.7 (f$_{\nu}$~$\propto$~$\nu^{-0.7}$) was assumed, to convert the OM UV flux at 2120~\AA\ (observer's frame; for UM~269, the V-band flux at 5500~\AA\ was used) to an estimate of the value at 2500~\AA\ (in the rest frame), needed for the definition of $\alpha_{ox}$, which is:

\begin{equation}
\alpha_{ox} = \frac{log\frac{f_{\nu}(2 keV)}{f_{\nu}(2500\AA)}}{log\frac{{\nu}(2 keV)}{{\nu}(2500 \AA)}}
\label{alphaoxeqn}
\end{equation}

\begin{table*}
\begin{center}
\caption{Optical and UV magnitudes obtained from the OM. The corresponding wavelengths for the different bands are: V -- 550~nm; B -- 440~nm; U -- 360~nm; UVW1 -- 291~nm; UVM2 -- 231~nm; UVW2 -- 212~nm. The last column gives the calculated two-point optical to X-ray slope, $\alpha_{ox}$. $^{a}$ The V-band magnitude for UM 269 has been taken from the Hewitt \& Burbidge catalogue (Hewitt \& Burbidge 1993), since no OM data were obtained.} 
\label{omdata}
\begin{tabular}{p{1.9truecm}p{1.0truecm}p{1.0truecm}p{1.0truecm}p{1.0truecm}p{1.0truecm}p{1.0truecm}p{1.0truecm}}
\hline
 & \multicolumn{6}{c}{mean optical and UV magnitudes}\\
Object & V & B & U & UVW1 & UVM2 & UVW2 & $\alpha_{ox}$\\
\hline
Q~0144$-$3938 & --- & 16.7 & 15.6 & 15.5 & ---& 15.1 & $-$1.47\\
UM 269 &  17.85$^{a}$ & --- & --- & --- & --- & --- & $-$1.40\\
PG~1634+706 & 14.8 & --- & 14.1 &--- & --- & 14.1 &  $-$1.57\\
SBS~0909+532 & ---&--- & ---& 15.4 & 15.8 & 15.8 &  $-$1.44\\
PG~1247+267 & ---&--- &--- & 15.0 & 15.3 &  15.6 & $-$1.55\\

\hline
\end{tabular}
\end{center}
\end{table*}

\begin{figure*}
\begin{center}
\includegraphics[clip,width=5.0cm,angle=-90]{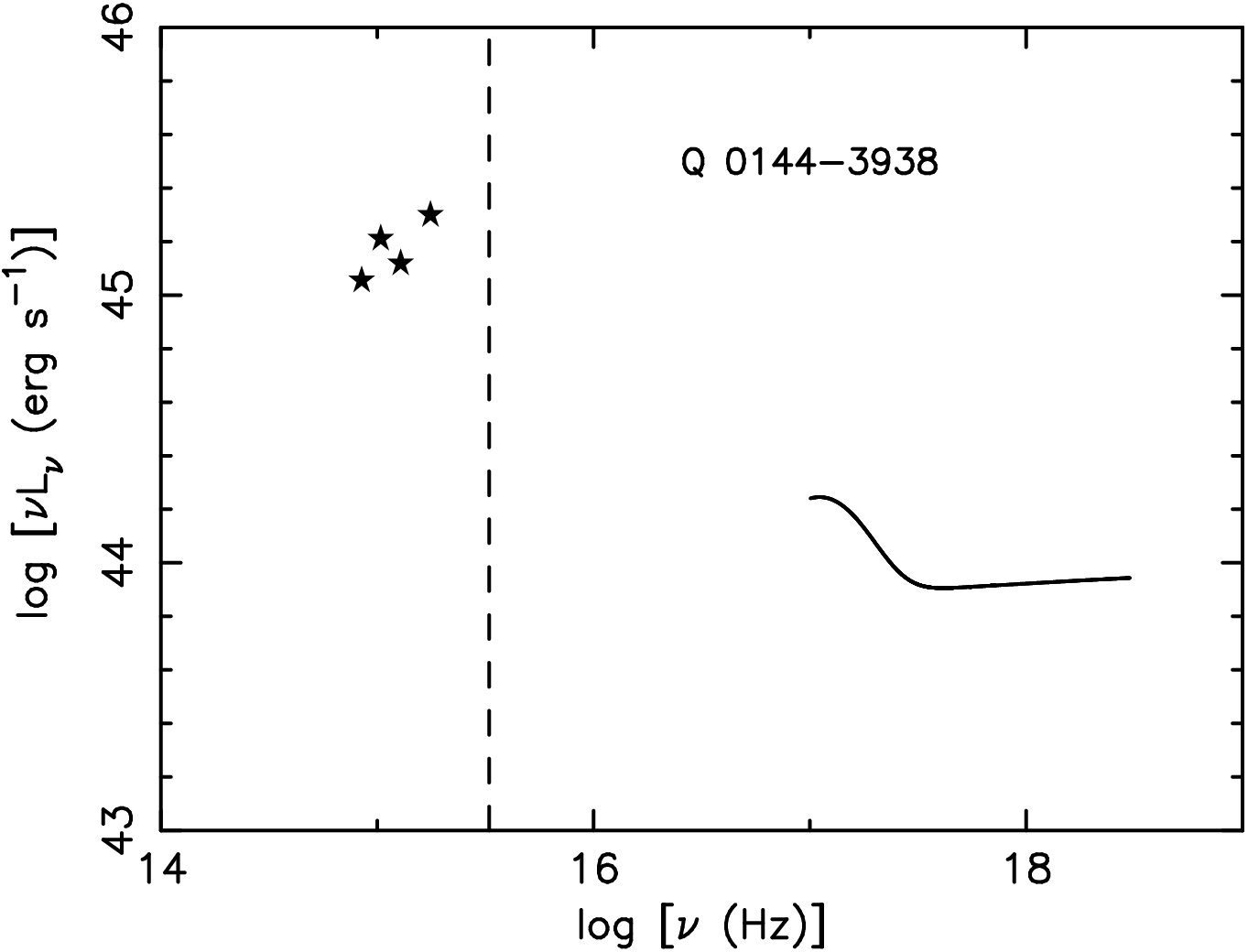}\hspace*{0.5cm}
\includegraphics[clip,width=5.0cm,angle=-90]{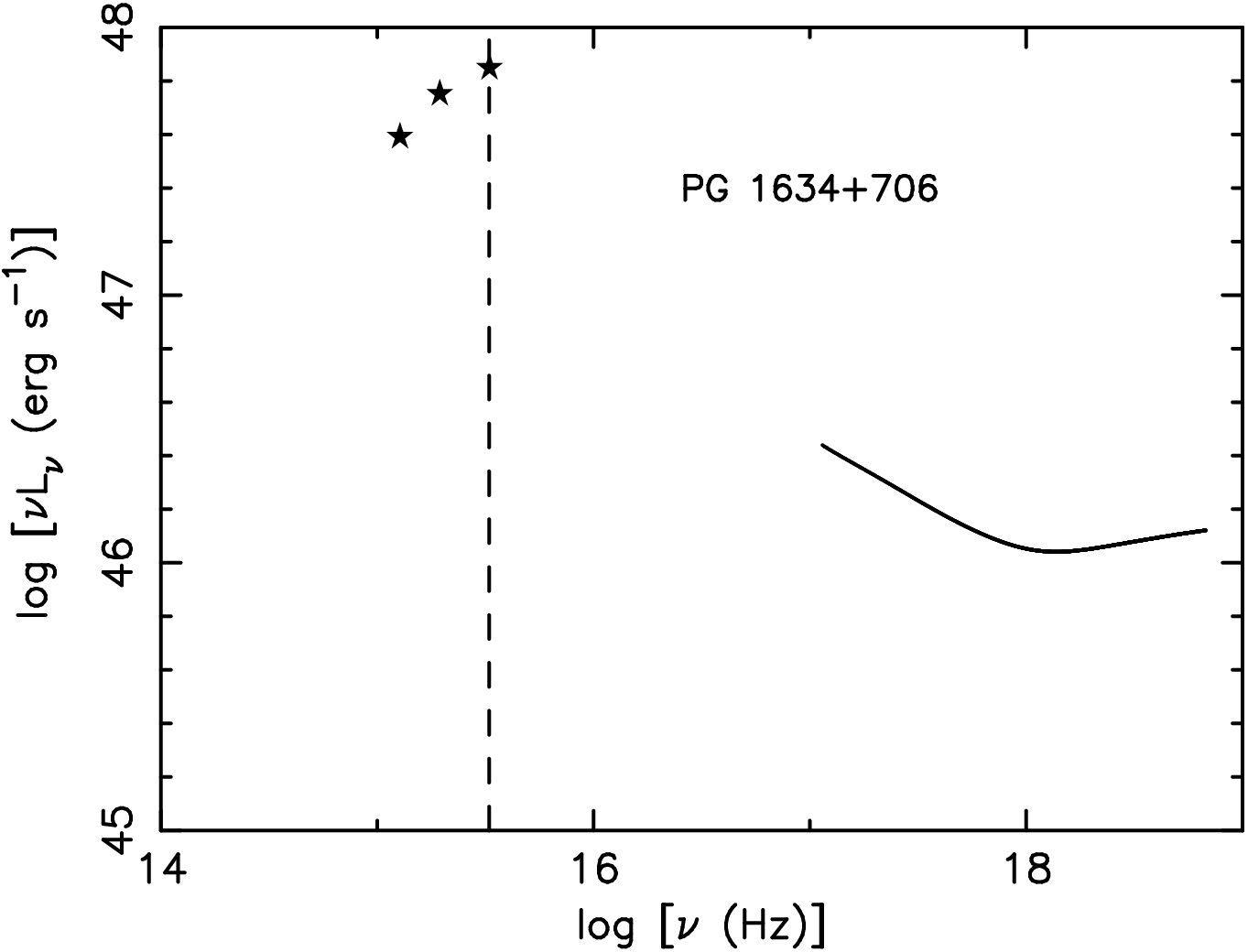}\vspace*{0.7cm}
\includegraphics[clip,width=5.0cm,angle=-90]{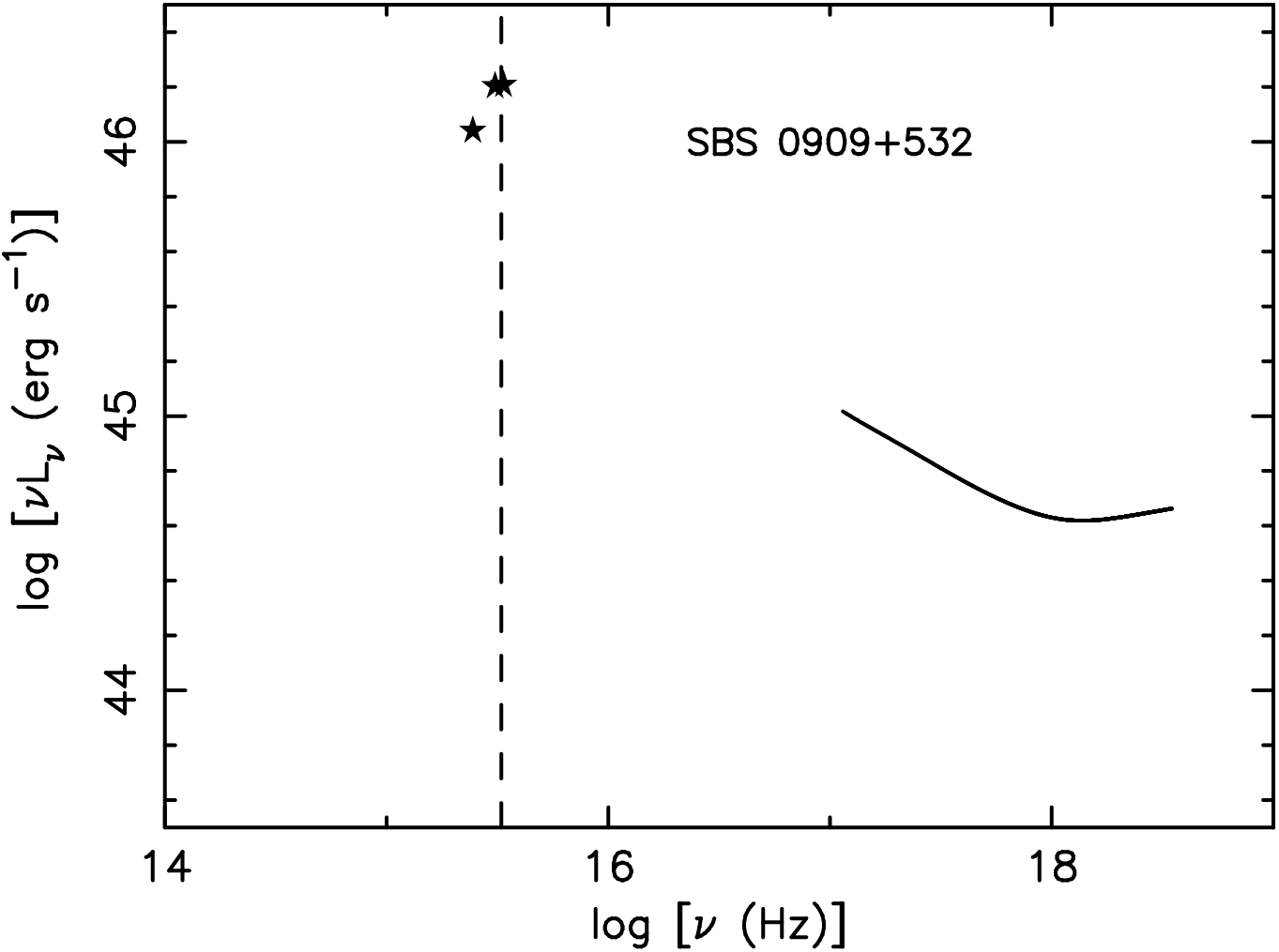}\hspace*{0.5cm}
\includegraphics[clip,width=5.0cm,angle=-90]{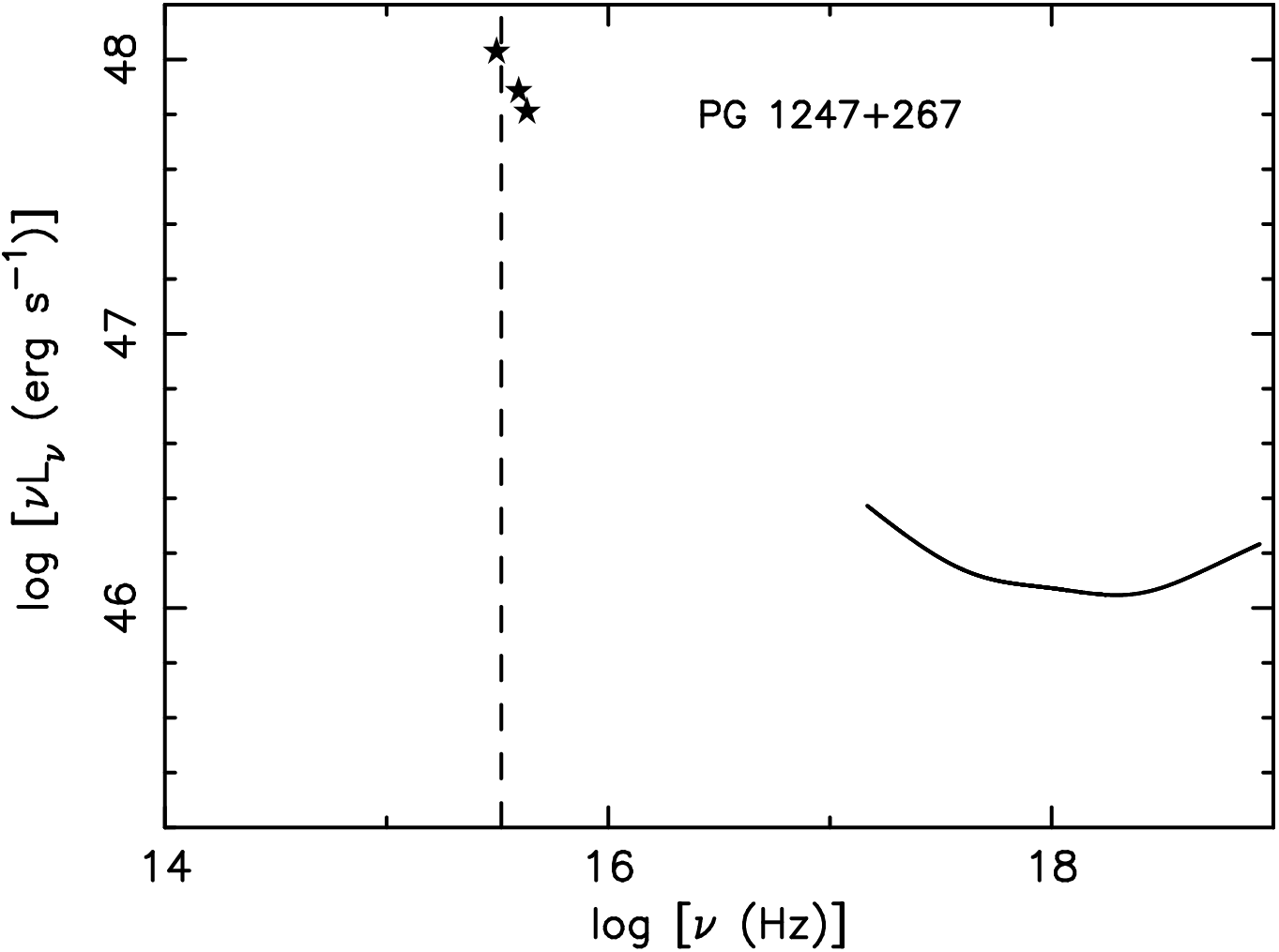}
\caption{Spectral Energy Distributions, plotting the EPIC and (dereddened) OM data. The vertical dotted line signifies the Lyman limit (912 \AA). The abscissa gives the frequency in the rest frame. The EPIC data were modelled with two {\it thComp} components, and the unabsorbed data have been plotted. The OM points (shown as stars) have been dereddened and corrected for Lyman line absorption. Note, however, that the two points at higher frequencies than the Lyman limit in PG~1247+267 have {\it only} been corrected for absorption due to the Lyman lines, {\it not} the Lyman continuum as well; they are, therefore, too red.}
\label{sed}  
\end{center}
\end{figure*}

SEDs (Spectral Energy Distributions) were produced, showing the Comptonization fit over the {\it XMM-Newton} band, together with the OM optical/UV points, and are plotted in Figure~\ref{sed}. There is an added complication for the objects at higher redshift, however. For PG~1634+706, SBS~0909+532 and PG~1247+267, some of the Optical Monitor data points are very close to the Lyman limit of 912~\AA; a vertical dotted line indicates 912~\AA\ in Figure~\ref{sed}. Without applying a correction factor, the optical/UV data points would be seen to fall with increasing frequency, since some of the higher-frequency photons have been absorbed. Barvainis (1990) plots an SED for PG~1634+706 at frequencies lower than $\sim$~2~$\times$~10$^{15}$~Hz, where the spectrum can be seen to be rising as the frequency increases. Similarly, Koratkar, Kinney \& Bohlin (1992) show a UV spectrum of PG~1247+267, where absorption below $\sim$~2000~\AA\ can clearly be seen. The measurements which fall between Ly$\alpha$ (1216~\AA) and Ly$\beta$ (1026~\AA), or between Ly$\beta$ and the limit of the series, have been corrected by the mean values for the Lyman line absorption given by O'Brien, Gondhalekar \& Wilson (1988). Beyond the Lyman limit, absorption by the continuum is also important. This correction factor is, however, uncertain, so the two highest frequency UV points in PG~1247+267 have only been corrected for the line absorption, not the additional continuum absorption; therefore, the UV spectrum in this case is reddened.




\section{Discussion}
\label{disc}

In this paper, optical to X-ray SEDs have been investigated for high luminosity, radio-quiet QSOs, ranging from $\sim$~7~$\times$~10$^{44}$ to 3~$\times$~10$^{46}$ erg s$^{-1}$ over the X-ray band (7~$\times$~10$^{45}$ -- 2~$\times$~10$^{48}$erg s$^{-1}$ bolometric).

When fitting the rest-frame 2--10~keV band, the mean photon index for the five QSOs in this sample is $\Gamma$~=~1.93~$\pm$~0.02. Despite the sample being small, this value is in very good agreement with the slopes measured in {\it ASCA} spectra by Reeves \& Turner (2000), for which a mean slope of 1.89~$\pm$~0.05 was found for the radio-quiet QSOs, and also with {\it Ginga} results (Lawson \& Turner 1997). Similarly, Page \etal (2003a) also obtained an average value of 1.89~$\pm$~0.04, for a sample of lower luminosity serendipitous {\it XMM-Newton}-detected QSOs. This value of $\Gamma$~$\sim$~1.9 is the same as that found for Seyfert 1 galaxies by Nandra \& Pounds (1994). The mean $\alpha_{ox}$ for this sample was found to be $-$1.48. This is somewhat flatter than other surveys of QSOs have found (e.g., 1.66 -- Page \etal 2003a; 1.56--1.78, Vignali \etal 2001), but the result is  not surprising, since the five objects in this sample were selected from luminous AGN in the {\it ROSAT} All Sky Survey and are, therefore, X-ray bright. $\alpha_{ox}$ has been previously found to become steeper with increased luminosity (e.g., Wilkes \etal 1994; Vignali, Brandt \& Schneider 2003). This small sample confirms these results, with a simple Spearman Rank correlations giving a probability of 98~per~cent.


The vast majority of luminous AGN (Turner \& Pounds 1989; Pounds \& Reeves 2002) show soft excesses -- that is, emission above the extrapolation of a power-law, fitted to the 2--10 keV (rest frame) energy band, at energies $\lo$~1~keV. Four of the QSOs in this sample show soft excess emission very clearly; the evidence is only marginal for PG~1247+267, where the reflection component provides the observed curvature in the spectrum at higher energies. It should be noted that, at higher redshifts, cool soft excesses will become more difficult to detect, due to the shift in the rest-frame energy-band to higher values. PG~1247+267 shows a weak indication for a blackbody component of kT~$\sim$~50~eV, which, at z~=~2.038, is mainly shifted out of the bandpass of {\it XMM-Newton}. Alternatively, if the spectrum is modelled without a reflection parameter, then PG~1247+267 shows a very hot soft excess (highest blackbody kT of $\sim$~0.7~keV); this lack of reflection is not consistent with the strong iron line, though, unless the reprocessing material is Compton-thin.. 

Thus, it appears that high-luminosity radio-quiet AGN have identical X-ray continuum properties to lower-luminosity QSOs and Seyfert 1 galaxies. The objects in this sample cover a range of X-ray luminosities between $\sim$~7~$\times$~10$^{44}$ and 3~$\times$~10$^{46}$ erg~s$^{-1}$, whereas Seyfert galaxies typically cover a range of $\sim$~10$^{42}$--10$^{44}$ erg~s$^{-1}$ (Nandra \& Pounds 1994). This indicates that the AGN X-ray continuum shape remains essentially constant over a wide range of black hole mass and luminosity. 

Two components are required to model the broad-band X-ray continuum. These can be explained through Comptonization of photons from the accretion disc: the hard X-ray power-law is produced through an interaction with hot (possibly non-thermal) electrons in a corona above the accretion disc. A second, cooler population of electrons ($\sim$~0.5~keV, rather than $>$~100~keV) produces the soft excess; these electrons are still much hotter than the accretion disc itself, though, which has a temperature of kT~$\lo$~30~eV. It is not yet clear how the soft excess electrons are heated. One possibility is that some areas of the accretion disc are irradiated by the hot electron corona; this could lead to `hot-spots' on the accretion disc, where a warm, mildly ionized layer could Comptonize the thermal optical/UV photons to the soft excess temperatures (e.g., di Salvo \etal 2001). Alternatively, the regions of the disc heated by the flares could thermalise the radiation, forming a blackbody spectrum with a temperature higher than that of the mean disc emission (Haardt, Maraschi \& Ghisellini 1994). The size of this soft-excess-forming `region' can be estimated by assuming that the emission is blackbody in nature (since $\tau$~$>>$~1, this is a valid approximation); taking the emitting region to be circular, and performing this calculation, leads to a radius of $\sim$10$^{11}$--10$^{12}$~m. For a black hole of mass 10$^{9}$~$\msun$, the Schwarzschild radius is $\sim$~3~$\times$~10$^{12}$~m -- i.e., the emitting region is actually smaller than R$_{sch}$ This implies that the warmer region is patchy, rather than covering the entire surface of the disc; likewise, the hot, flaring (coronal) regions must also be small. This result is also consistent with some of the disc photons being observed directly as the optical/UV emission, while yet more interact with the hot electron corona to form the observed power-law.

Iron lines were commonly identified in AGN spectra by {\it Ginga} observations (Pounds \etal 1989; Pounds \etal 1990; Nandra \etal 1991; Nandra \& Pounds 1994). With the improved resolution of current instruments, it is sometimes possible to differentiate between narrow ($<$10~eV; unresolved by an instrument such as {\it XMM-Newton}) and broad lines. To date, few broad lines have been found in {\it XMM-Newton} spectra of AGN, examples being Q~0056$-$363 (Porquet \& Reeves 2003), MCG$-$5$-$23$-$16 (Dewangan, Griffiths \& Schurch 2003), MCG$-$6$-$30$-$15 (Fabian \etal 2002b),  Mrk~205 (Reeves \etal 2001) and Mrk~509 (Pounds \etal 2001). Nandra \etal (1997) reported a Baldwin effect for the broad lines in {\it ASCA} data -- that is, as the X-ray luminosity of the source increases, the strength of the broad line diminishes. The same Baldwin effect has been identified for the narrow iron lines in {\it XMM-Newton} observations (Page \etal 2004). Also, both Reeves \etal (1997) and Lawson \& Turner (1997) found that reprocessed features (such as the iron K$\alpha$ line and reflection humps) were less common in QSOs than in the lower-luminosity Seyferts.  Thus, it is rather surprising to find a strong, apparently broad line, together with a clear reflection component, in such a luminous QSO as PG~1247+267 (broad-band X-ray luminosity of $\sim$~2~$\times$~10$^{46}$ erg~s$^{-1}$). Both the strength and velocity width of the line ($\sigma$~=~30~000 km~s$^{-1}$) imply that the reflection is occuring off the inner accretion disc, rather than distant matter, such as the torus. The line is also much stronger than the narrow component of the Fe lines observed from further out (that is, from the Broad or Narrow Line Regions, or the torus) in many Seyfert 1s; these narrow lines are typically around 100~eV in equivalent width (e.g., Page \etal 2004; Kaspi \etal 2001; Pounds \etal 2001; Reeves \etal 2001). Q~0144$-$3938 was also better fitted with a broad, rather than a narrow line, but is less luminous than PG~1247+267; it was not possible to determine whether the line in the spectrum of SBS~0909+267 was broad or narrow. 3C~273, a lumious radio-loud quasar, has previously been reported to show evidence for a broad line (Yaqoob \& Serlemitsos 2000; Page \etal 2004b), though much weaker than in PG~1247+267, with an EW of $\sim$~50~eV. PG~1247+267 seems to be the most luminous QSO to show a strong (apparently) broad line to date, although there is some uncertainty about whether the line is truly broad; the object also reveals the first clear detection of a reflection component in a high-luminosity AGN. 

Reeves \& Turner (2000) found, in a sample of 62 QSOs (both radio-loud and -quiet), that approximately half of the iron emission lines seen were at energies $>$6.4~keV, implying the lines originated from partially ionized matter. It seems somewhat unusual that the lines reported in this paper appear neutral; however, the current sample of five (three detections) is much smaller than in the Reeves \& Turner (2000) paper.

It should be noted that the reflection component found in PG~1247+267 is very strong, even for an accretion disc: R~$>$~2 implies the reflection is occuring from $>$~4$\pi$ steradian. There are a number of possible explanations for this. It could be a geometrical phenomenon; i.e., part of the direct, hard X-ray emission may be being obscured by structure within the disc. In this case, the reflection would appear enhanced (see, e.g., Fabian \etal 2002a). Alternatively, it could be due to relativistic effects. Fabian \& Vaughan (2003) use gravitational light-bending as an argument for explaining the very strong line (also $\sim$~400-500~eV) observed in MCG~$-$6$-$30$-$15.

With the discovery of such a strong iron line and reflection component in PG~1247+267, a much longer observation of the object ($>$19~ks of PN time as presented here) is desirable, in order to investigate the spectral features more thoroughly, including whether the line varies over time.

\section{ACKNOWLEDGMENTS}
The work in this paper is based on observations with {\it
XMM-Newton}, an ESA
science mission, with instruments and contributions directly funded by
ESA and NASA. The authors would like to thank the EPIC Consortium for all their work during the calibration phase, 
and the SOC and SSC teams for making the observation and analysis
possible. 
This research has made use of the NASA/IPAC Extragalactic
Database (NED), which is operated by the Jet Propulsion Laboratory,
California Institute of Technology, under contract with the National
Aeronautics and Space Administation.


\begin{thebibliography}{}

\bibitem[Barvainis 1990]{ba90}Barvainis R., 1990, ApJ, 353, 419
\bibitem[Barvainis 1996]{ba96}Barvainis R., Lonsdale C., Antonucci R., 1996, AJ, 111, 1431
\bibitem[Becker 1995]{be95}Becker R.H., White R.L., Helfand D.J., 1995, ApJ, 450, 559 
\bibitem[Chartas 2000]{ch00}Chartas G., 2000, ApJ, 531, 81
\bibitem[Condon 1998]{co98}Condon J.J., Cotton W.D., Greisen E.W., Yin Q.F., Perley R.A., Taylor G.B., Broderick J.J., 1998, AJ, 115, 1693
\bibitem[Dewangan 2003]{de03}Dewangan G.C., Griffiths R.E., Schurch N.J., 2003, ApJ, 592, 52
\bibitem[Dickey 1990]{di90}Dickey J.M., Lockman F.J., 1990, ARA\&A, 28, 215
\bibitem[DiSalvo 2001]{dis01}di Salvo T., Done C., {\. Z}ycki P.T., Burderi L., Robba N.R., 2001, ApJ, 547, 1024 
\bibitem[Elvis 1994]{el94}Elvis M., Wilkes B.J., McDowell J.C., Green R.F., Bechtold J., Willner S.P., Oey M.S., Polomski E., Cutri R., 1994, ApJS, 95, 1

\bibitem[Fabian 2003]{fa03}Fabian A.C., Vaughan S., 2003, MNRAS, 340, L28
\bibitem[Fabian 2002]{fa021}Fabian A.C., Ballantyne D.R., Merloni A., Vaughan S., Iwasawa K., Boller Th., 2002a, MNRAS, 331, L35 
\bibitem[Fabian 2002]{fa02b}Fabian A.C. \etal, 2002b, MNRAS, 335, L1 
\bibitem[George 1991]{ge91}George I.M., Fabian A.C., 1991, MNRAS, 249, 352
\bibitem[George 2000]{ge00}George I.M., Turner T.J., Yaqoob T., Netzer H., Laor A., Mushotzky R.F., Nandra K., Takahashi T., 2000, ApJ, 531, 52
\bibitem[Haardt 1994]{ha94}Haardt F., Maraschi L., Ghisellin G., ApJ, 1994, 432, L95
\bibitem[Hewitt 1993]{hew93}Hewitt A., Burbidge G., 1993, ApJS, 87, 451
\bibitem[Kaspi 2001]{kas01}Kaspi S. \etal, 2001, ApJ, 554, 216 
\bibitem[Kellerman 1989]{kel89}Kellerman K.I., Sramek R., Schmidt M., Shaffer D.B., Green R., 1989, AJ, 98, 1195
\bibitem[Kochanek 1997]{ko97}Kochanek C.S., Falco E.E., Schild R., Dobrzycki A., Engels D., Hagen H.-J., 1997, ApJ, 479, 678
\bibitem[Koratkar 1992]{ko92}Koratkar A.P., Kinney A.L., Bohlin R.C., 1992, ApJ, 400, 435
\bibitem[Kukula 1998]{ku98}Kukula M.J., Dunlop J.S., Hughes D.H., Rawlings S., 1998, MNRAS, 297, 366
\bibitem[Lawson 1997]{la97}Lawson A.J., Turner M.J.L., 1997, MNRAS, 288, 920
\bibitem[Lehar 2002]{le02}Leh{\' a}r J. \etal, 2002, ApJ, 536, 584
\bibitem[Lightman 1988]{li88}Lightman A.P., White T.R., 1988, ApJ, 335, 57 
\bibitem[Magdziarz 1995]{mag95}Magdziarz P., Zdziarski A.A., 1995, MNRAS, 273, 837
\bibitem[Makishima 1986]{mak86}Makishima K., Maejima Y., Mitsuda K., Bradt H.V., Remillard R.A., Tuohy I.R., Hoshi R., Nakagawa M., 1986, ApJ, 308, 635
\bibitem[Mitsuda 1984]{mit84}Mitsuda K. \etal, 1984, PASJ, 36, 741
\bibitem[Nandra 1994]{nan94}Nandra K., Pounds K.A., 1994, MNRAS, 268, 405
\bibitem[Nandra 1997]{nan97}Nandra K., George I.M., Mushotzky R.F., Turner T.J., Yaqoob T., 1997, ApJ, 488, L91
\bibitem[Nandra 1991]{nan91}Nandra K., Pounds K.A., Stewart G.C., George I.M., Hayashida K., Makino F., Ohashi T., 1991, MNRAS, 248, 760
\bibitem[O'Brien 1988]{ob88}O'Brien P.T., Gondhalekar P.M., Wilson R., 1988, MNRAS, 233, 801
\bibitem[Oscoz 1997]{os97}Oscoz A., Serra-Ricart M., Mediavilla E., Buitrago J., Goicoechea L.J., 1997, ApJ, 491, L7
\bibitem[Page 2003]{pa03a}Page K.L., Turner M.J.L., Reeves J.N., O'Brien P.T., Sembay S., 2003a, MNRAS, 338, 1004
\bibitem[Page 2004]{pa04}Page K.L., O'Brien P.T., Reeves J.N., Turner M.J.L., 2004, MNRAS, 347, 316 
\bibitem[Page 2003]{pa04b}Page K.L., Turner M.J.L., Done C., O'Brien P.T., Reeves J.N., Sembay S., Stuhlinger M., 2004b, MNRAS, 349, 57
\bibitem[Peterson 1997]{pe97}Peterson B.M., 1997, An Introduction to Active Galactic Nuclei, CUP, Cambridge, UK, ISBN 0521479118
\bibitem[Porquet 2003]{por03}Porquet D., Reeves J.N., 2003, A\&A, 408, 119
\bibitem[Pounds 2002]{po02}Pounds K.A., Reeves J.N., 2002, in New Visions of the X-ray Universe in the {\it XMM-Newton} and {\it Chandra} era (astro-ph/0201436)
\bibitem[Pounds 2001]{po01}Pounds K., Reeves J., O'Brien P., Page K., Turner M., Nayakshin S., 2001, 559, 181
\bibitem[Pounds 1990]{po90}Pounds K.A., Nandra K., Stewart G.C., George I.M., Fabian A.C., 1990, Nature, 344, 132
\bibitem[Pounds 1989]{po89}Pounds K.A., Nandra K., Stewart G.C., Leighly K., 1989, MNRAS, 240, 769
\bibitem[Protassov 2002]{pr02}Protassov R., van Dyk D.A., Connors A., Kashyap V.L., Siemiginowska A., 2002, ApJ, 571, 545
\bibitem[Reeves 2003]{re03b}Reeves J.N., O'Brien P.T., Ward M.J., 2003, ApJ, 593, L65
\bibitem[Reeves 2001]{re01}Reeves J.N., Turner M.J.L., Pounds K.A., O'Brien P.T., Boller Th., Ferrando P., Kendziorra E., Vercellone S., 2001, A\&A, 365, L134
\bibitem[Reeves 2000]{re00}Reeves J.N., Turner M.J.L., 2000, MNRAS, 316, 234
\bibitem[Reeves 1997]{re97}Reeves J.N., Turner M.J.L., Ohashi T., Kii T., 1997, MNRAS, 292, 468
\bibitem[Turner 1989]{tu89}Turner T.J., Pounds K.A., 1989, MNRAS, 240, 833
\bibitem[Vignali 2003]{vi03}Vignali C., Brandt W.N., Schneider D.P., 2003, AJ, 125, 433
\bibitem[Vignali 1999]{vi99}Vignali C., Comastri A., Cappi M., Palumbo G.G.C., Matsuoka M., Kubo H., 1999, ApJ, 516, 582
\bibitem[Wadadekar 1999]{wa99}Wadadekar Y., Kembhavi A., 1999, AJ, 118, 1435
\bibitem[Wilkes 1987]{wi87}Wilkes B.J., Elvis M., 1987, ApJ, 323, 243
\bibitem[Wilkes 1994]{wi94}Wilkes B.J., Tananbaum H., Worrall D.M., Avni Y., Oey M.S., Flanagan J., 1994, ApJS, 92, 53
\bibitem[Worrall 1987]{wo87}Worrall D.M., Giommi P., Tananbaum H., Zamorani G., 1987, ApJ, 313, 596
\bibitem[Yaqoob 2000]{ya00}Yaqoob T., Serlemitsos P., 2000, ApJ, 544, L95
\bibitem[Yuan 1998]{yu98}Yuan W., Brinkmann W., Siebert J., Voges W., 1998, A\&A, 330, 108
\bibitem[Zamorani 1981]{za81}Zamorani G. \etal, 1981, ApJ, 245, 357
\bibitem[Zdziarski 1996]{zd96}Zdziarski A.A., Johnson W.N., Magdziarz P., 1996, MNRAS, 283, 193

\end{thebibliography}
\end{document}